\documentclass[10pt]{iopart}

\usepackage{iopams} 
\usepackage{graphicx}
\usepackage{bm}
\usepackage{cite}
\begin{document}

\title[]{Spatial distribution of thermoelectric voltages in a Hall-bar shaped two-dimensional electron system under a magnetic field}

\author{Akira Endo, Kazuhiro Fujita, Shingo Katsumoto and Yasuhiro Iye}

\address{Institute for Solid State Physics, The University of Tokyo, Kashiwanoha, Kashiwa, Chiba 277-8581, Japan}
\ead{akrendo@issp.u-tokyo.ac.jp}
\vspace{10pt}
%\begin{indented}
%\item[]February 2014
%\end{indented}

\begin{abstract}
We have investigated the spatial distribution of the electron temperature generated in a two-dimensional electron system (2DES) subjected to a perpendicular magnetic field. 
%By introducing a temperature gradient employing the current-heating technique, we
We measure thermoelectric voltages between Ohmic contacts located at the end of the voltage-probe arms of a Hall bar fabricated from a GaAs/AlGaAs 2DES wafer, immersed in the mixing chamber of a dilution refrigerator held at 20 mK\@. Magneto-oscillations due to the Landau quantization are examined for the thermoelectric voltages between the contact pairs straddling the main bar (arrangement to measure the transverse component $V_{yx}$), and between the pairs located along the same side of the main bar (arrangement for the longitudinal component $V_{xx}$). For the former arrangement, the oscillation amplitude diminishes with the distance from the heater. For the latter arrangement, the pair on one side exhibits much larger amplitude than the pair on the opposite side, and the relation becomes reversed by inverting the magnetic field. The behaviours of the oscillation amplitude are qualitatively explained by the spatial distribution of the electron temperature numerically calculated taking into consideration the thermal diffusion into the voltage contacts and the electron-phonon interaction. For both arrangements, the oscillations are shown to derive predominantly from the transverse (Nernst) component, $S_{yx}$, of the thermopower tensor. The calculation also reveals that the voltage probes, introducing only minor disturbance at zero magnetic field, substantially reduce the temperature once a magnetic field is applied, and the thermoelectric voltages generated at the voltage arms account for a significant part of the measured voltages.
\end{abstract}

% \noindent{\it Keywords}: thermoelectric voltage, two-dimensional electron system, electron temperature, magnetic field, Hall bar, diffusion contribution, quatnum oscillations

%\submitto{\JPCM}
%\maketitle
\ioptwocol

\section{Introduction}

A large Hall angle approaching $\pi/2$ is one of the key parameters characterizing a high-mobility two-dimensional electron system (2DES) subjected to a perpendicular magnetic field. The electron density $n_\mathrm{e}$ in a 2DES residing at a GaAs/AlGaAs heterointerface is typically $\sim$5$\times$10$^{15}$ m$^{-2}$ or less. There the Hall resistivity $\rho_{xy} \sim B/(n_\mathrm{e}e)$ far exceeding the diagonal resistivity $\rho_{xx}$ can be achieved with a relatively small magnetic field ($\sim$0.1 T), resulting in a large Hall angle $\delta = \arctan (\rho_{xy}/\rho_{xx})$. (We assume throughout the paper that the 2DES is isotropic. Therefore we have $\rho_{xx} = \rho_{yy}$, $\rho_{yx} = -\rho_{xy}$ and the corresponding relations for the conductivity). Let us consider a rectangular sample of a 2DES with the source and drain electrodes attached at opposite ends. The spatial distributions of the electric field ${\bf E}$ and the current density ${\bf j}$ induced in such a sample by the application of a source-drain bias have been calculated employing  classical electromagnetism \cite{Wick54,Wakabayashi78,Rendell81,Neudecker87}. At zero magnetic field, both ${\bf E}$ and ${\bf j}$ are parallel to the longitudinal direction (the direction from the source to the drain). A magnetic field drastically alters the distributions of ${\bf E}$ and ${\bf j}$, primarily by introducing the angle $\delta$ between the two vectors. Rather counterintuitively, ${\bf E}$ becomes nearly perpendicular ($\sim \delta \simeq \pi/2$) to the longitudinal direction, barring the regions close to the electrodes. Small areas with highly concentrated ${\bf j}$ and the equipotential lines, dubbed ``hot spots'', emerge at the two diagonally opposite corners, one facing the source and the other facing the drain electrode. The predicted spatial distributions have been essentially verified by various experimental techniques \cite{Fontein92,Yacoby91,Ahlswede01,Kawano10JJAP}.

Replacing the source-drain bias with a temperature difference, we expect analogous spatial distributions in the temperature gradient $\boldsymbol{\nabla} T$ and the thermal flux density ${\bf j}_\mathrm{Q}$. Good analogy is expected when the electron and the lattice systems are virtually decoupled and major part of ${\bf j}_\mathrm{Q}$ is carried by electrons, as is the case in the GaAs/AlGaAs 2DES at very low temperatures (below $\sim$200 mK). The distribution of $\boldsymbol{\nabla} T$, especially the reorientation of $\boldsymbol{\nabla} T$ by a magnetic field, can profoundly affect the thermoelectric voltage generated by the temperature difference. In the analysis of the thermoelectric voltages of a 2DES, however, $\boldsymbol{\nabla} T$ is usually assumed to simply remain parallel to the longitudinal direction even in the presence of the magnetic field. Strangely, the redistribution of $\boldsymbol{\nabla} T$ has not attracted due attention it deserves. (The effect of the redistribution of $\boldsymbol{\nabla} T$ on the thermoelectric voltages was previously considered for bulk BiSb and InSb alloys \cite{Ertl63,Okumura98,Heremans00,Heremans01}.)

The purpose of the present study is to interpret experimentally measured thermoelectric voltages in terms of $\boldsymbol{\nabla} T$ calculated considering the effect of the magnetic field. Measurements were performed on a 2DES having a Hall-bar geometry. We measure voltages between several pairs of voltage probes and examine how the voltage varies with varying location on the Hall bar, focusing on the amplitude of the quantum oscillations due to the Landau quantization. An important difference between ${\bf j}$ and ${\bf j}_\mathrm{Q}$ in Hall-bar measurements is that the latter is allowed to flow into the voltage contacts, while the former is prohibited. Therefore, the arms of the voltage probes can host non-vanishing $\boldsymbol{\nabla} T$ and thus can contribute to the thermoelectric voltages \cite{Fletcher86}. We numerically calculate the spatial distribution of the electron temperature $T$, taking into account the thermal diffusion through the arms into the contacts as well as the power transferred to the lattice via electron-phonon interaction. The lattice temperature is assumed to be uniform and kept at the lowest temperature in the system. We find that the dependence of the measured thermoelectric voltages on the location and on the magnetic field can be qualitatively explained by the calculated spatial distribution of $T$. The calculations highlight crucial roles played by the voltage probes in reducing the temperature and in generating the thermoelectric voltage, when placed in a magnetic field.
The geometry of the voltage-probe arms designed to minimize the thermal disturbance turns out to be effective only at around zero magnetic field.
%The arms of the voltage probes are designed to minimize thermal disturbance employing a very low aspect ratio (the width divided by the length), and the design basically works at zero magnetic field. Interestingly, however, a magnetic field disables the capability of the low aspect ratio to suppress the thermal flux, letting the voltage probes to substantially affect the temperature.

Thermoelectric voltages are sensitive to the energy dependence of the conductivity and also are a measure of the entropy of the system \cite{Harman67,Callen85,Pottier10}, and thus have been extensively applied to the studies of transport properties and scattering mechanisms. Numbers of studies have been performed on the thermoelectric properties of a 2DES (see \cite{Gallagher92,Fletcher99} for reviews). Recent theoretical suggestion of the possibility to probe, through the entropy, non-Abelian quasiparticles in $\nu = 5/2$ fractional quantum Hall state \cite{Yang09,Barlas12} revitalized the interest in the thermoelectric voltages of a 2DES subjected to a magnetic field \cite{Chickering10,Chickering13,Kobayakawa13}. The present study suggests precautions to be taken in analysing the thermoelectric voltages measured in a magnetic field.

%\cite{Goswami09}

The paper is organized as follows. In section \ref{meas}, we describe experimental details and the results of the thermoelectric-voltage measurements. Section \ref{calcT} is devoted to the calculation of the spatial distribution of $T$. We start by briefly describing simple one-dimensional and rectangular models neglecting the electron-phonon interaction. Analytic solutions given for these models help us grasp the essence of the role played by a magnetic field. Numerical calculations follow, with which we examine the effect of the voltage probes and of the electron-phonon interaction. Thermoelectric voltages resulting from the distribution of $T$ thus calculated are presented in section \ref{thermvol}. We use thermoelectric coefficients deduced relying on the generalized Mott's formula. The calculated thermoelectric voltages are compared with the measurements presented in section \ref{meas}. We also evaluate separately the contribution of the main bar and of the voltage-probe arms to the thermoelectric voltages. Discussion on quantitative disparities between measured and calculated thermoelectric voltages, as well as on the characteristics and the limitations of the measurements in the present study, is given in section \ref{discussion}. Section \ref{conclusion} concludes the paper.

\section{Measurements \label{meas}}
\begin{figure*}[tb]
\begin{center}
\includegraphics[width=11cm,clip]{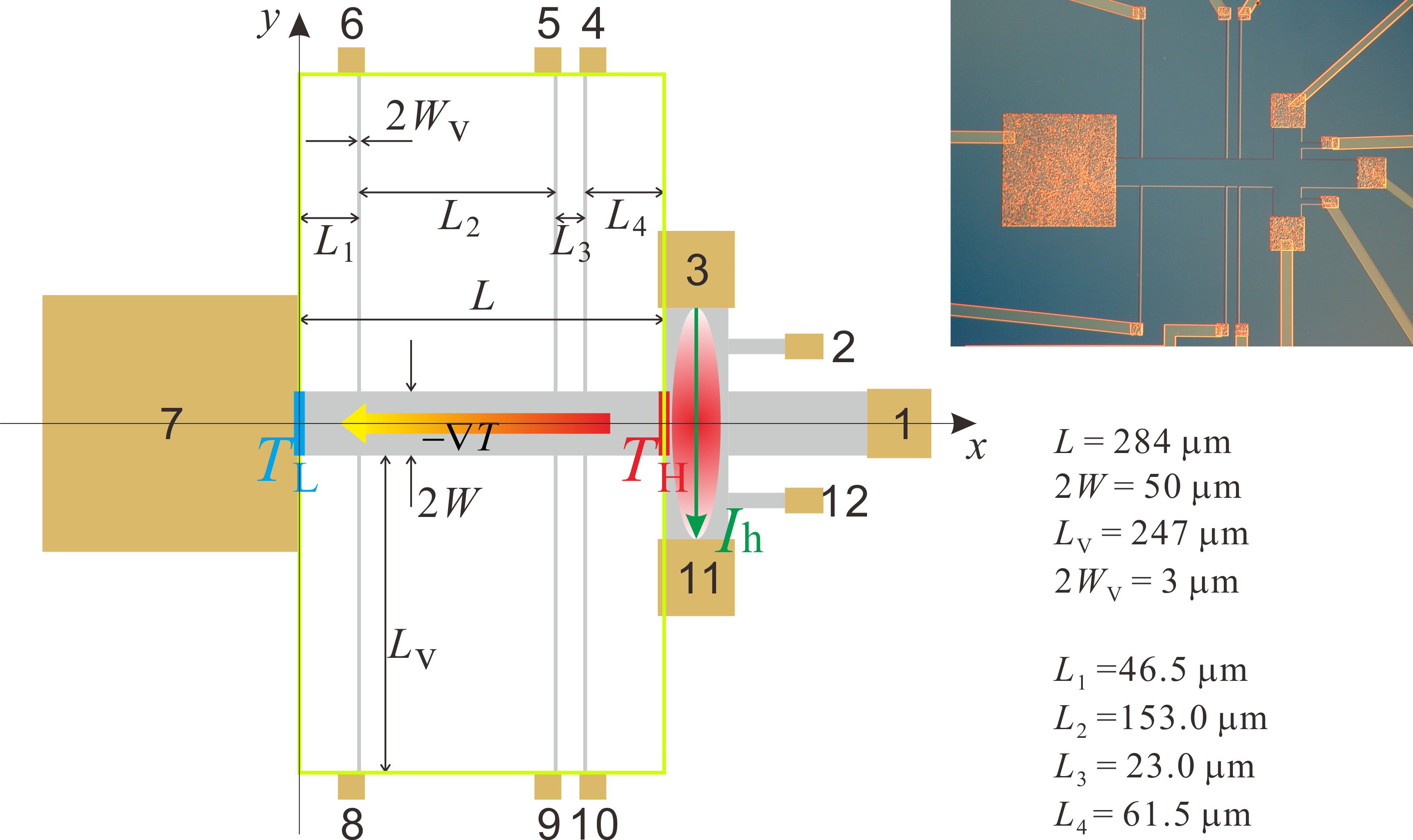} 
\caption{Schematic diagram of the Hall-bar device used in the present study. The main (horizontal) Hall bar (from 1 to 7, length of the effective part $L$, width $2W$) is fitted with six voltage probes (4, 5, 6, 8, 9, 10) to measure the thermoelectric voltages. The arms of the voltage probes are designed to be thin ($2W_\mathrm{V}$) and long ($L_\mathrm{V}$), in order to minimize the thermal disturbance. %to the main Hall bar. 
The secondary (vertical) Hall bar (from 3 to 11) serves as a heater and also contains two voltage probes (2 and 12) to monitor the electron temperature by the amplitude of the SdH oscillations. The light-green rectangle encompasses the main Hall bar, i.e., the main bar from the heater to the heat sink (7) and the arms of the voltage probes, for which the spatial distribution of the electron temperature is calculated in section \ref{calcTFEM} and plotted in figure \ref{TAephcalc}. Upper right inset: optical micrograph of the device. The dimensions of the Hall bar, including the locations of (the centres of) the arms, are listed to the right of the figure.}
\label{HallBarFig}
\end{center}
\end{figure*}

Thermoelectric-voltage measurements were performed on a Hall bar device depicted in figure \ref{HallBarFig}, fabricated from a GaAs/AlGaAs 2DES wafer having the electron density $n_\mathrm{e} = 3.7 \times 10^{15}$ m$^{-2}$ and the mobility $\mu = 80$ m$^2$V$^{-1}$s$^{-1}$. The device is composed of two crossing Hall bars \cite{Fujita10E}. The main (horizontal) Hall bar contains six voltage probes, allowing us to measure the transverse voltage $V_{yx}$ at three different locations along the main bar and the longitudinal voltage $V_{xx}$ both on the top and the bottom edges. The secondary (vertical) Hall bar is utilized as a heater to introduce the temperature gradient towards the other end (heat sink) of the main Hall bar. Joule heating by an ac (frequency $f$) heating current $I_\mathrm{h}$ raises the electron temperature in the secondary Hall bar, and the temperature is monitored by the amplitude of the Shubnikov-de Hass (SdH) oscillations. Since the Joule heating varies as $\propto {I_\mathrm{h}}^2$, the resulting thermoelectric voltage can be measured by detecting the component of $V_{yx}$ and $V_{xx}$ having the frequency $2f$, employing standard low-frequency ($f = 13$ Hz) ac lock-in technique. The measurements were carried out in a dilution refrigerator (Oxford, Kelvinox TLD) equipped with a superconducting magnet, with the sample immersed in the mixing chamber held at $T_\mathrm{bath} = 20$ mK\@.

Employing the current $I_h = 200$ nA, we obtain the electron temperature $T_\mathrm{H} = 330$ mK at the heater section. The current is chosen to be much larger than that used in measuring the resistivity (typically 0.5--10 nA), but kept small enough to avoid heating the lattice via the electron-phonon interaction. The gradient is thus introduced only into the electron temperature, leaving the lattice temperature, assumed to be at $T_\mathrm{bath}$, intact. This enables us to selectively measure the diffusion contribution \cite{Maximov04,Chickering09,Fujita10E,Goswami11} and eliminate the phonon-drag contribution. Note that the latter can, under certain circumstances (in a standard experimental method employing an external heater, which introduces the temperature gradient also into the lattice temperature), become very large and dominate the thermoelectric voltage in a 2DES embedded in a thick ($\sim$500 $\mu$m) semiconductor wafer \cite{Fletcher86,Fletcher99}. 
Although the phonon-drag contribution is expected to be relatively small in the temperature range of the present study, its elimination by the present experimental technique is advantageous in avoiding possible complication of the interpretation.
Ohmic contacts at the end of the main and the secondary Hall bars and of voltage-probe arms are composed of diffused AuGeNi alloy. The large area (200$\times$200 $\mu$m$^2$) of the contact at the heat sink is devised to ensure good thermal contact to the bath. By contrast, the arms of the voltage probes are designed to be thin and long and are terminated by the contact with a small area (21$\times$21 $\mu$m$^2$), in an attempt to minimize the thermal disturbance to the main bar. %by the voltage probes.  
In what follows, however, we assume for simplicity that all the Ohmic contacts, including the small contacts for the voltage probes, are at the same temperature $T_\mathrm{L} = T_\mathrm{bath}$.  The appropriateness of this assumption will be examined in section \ref{discMandC}.

\begin{figure}[tb]
\begin{center}
\includegraphics[bb=40 20 455 606,width=8.3cm,clip]{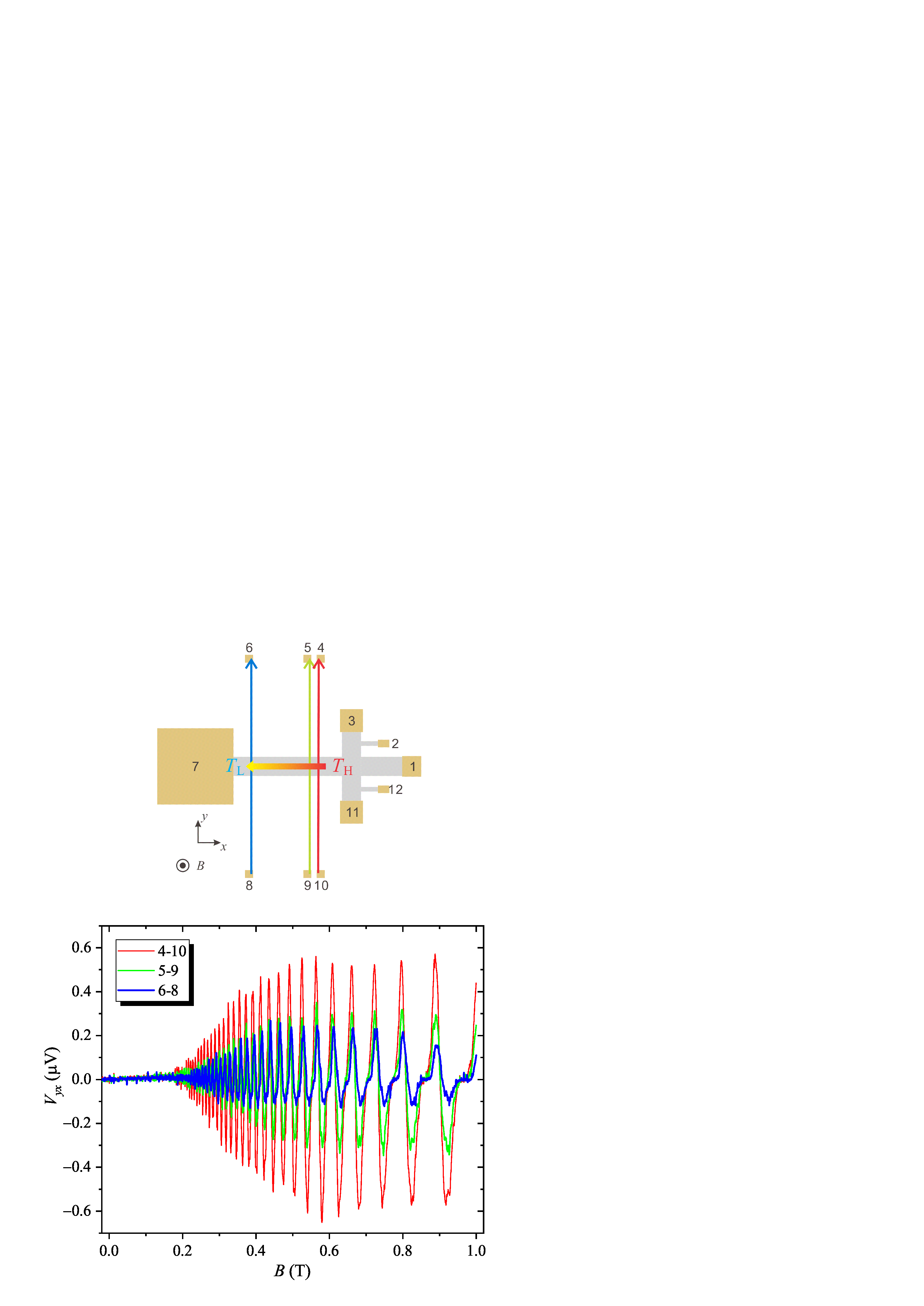} 
\caption{Magnetic-field dependence of the thermoelectric voltages $V_{yx}$ measured between the three pairs of probes, depicted by arrows in the top inset, located across the main Hall bar. (The direction of the arrows indicates the sign of $V_{yx}$.)}
\label{MeasVyx}
\end{center}
\end{figure}

The transverse thermoelectric voltages $V_{yx}$ measured with three pairs of voltage probes are plotted in figure \ref{MeasVyx} as a function of the magnetic field $B$. Quantum oscillations due to the Landau quantization (the equivalent of the SdH oscillations in the resistivity) commences at $B \simeq 0.2$ T\@. The amplitude increases with the increase of the magnetic field up to $B \simeq 0.5$ T, and then levels off and shows slight decrease with further increase of $B$. Above $\sim$0.55 T, the lineshape characteristic of the off-diagonal (Nernst) thermoelectric voltage in the quantum Hall systems  \cite{Jonson84} becomes apparent: namely, plateaus with $V_{yx} \sim 0$ at around integer fillings (only even integer fillings are resolved in this magnetic-field range) and the saw-tooth like lineshape with sign reversal in between (see $S_{yx}$ in figure S2 of the Supplementary data). The traces are antisymmetric with the reversal of the magnetic field. An important feature we want to stress here is that the oscillation amplitude becomes smaller for the voltage-probe pairs located farther away from the heater section.

\begin{figure}[tb]
\begin{center}
\includegraphics[bb=40 165 455 947,width=8.3cm,clip]{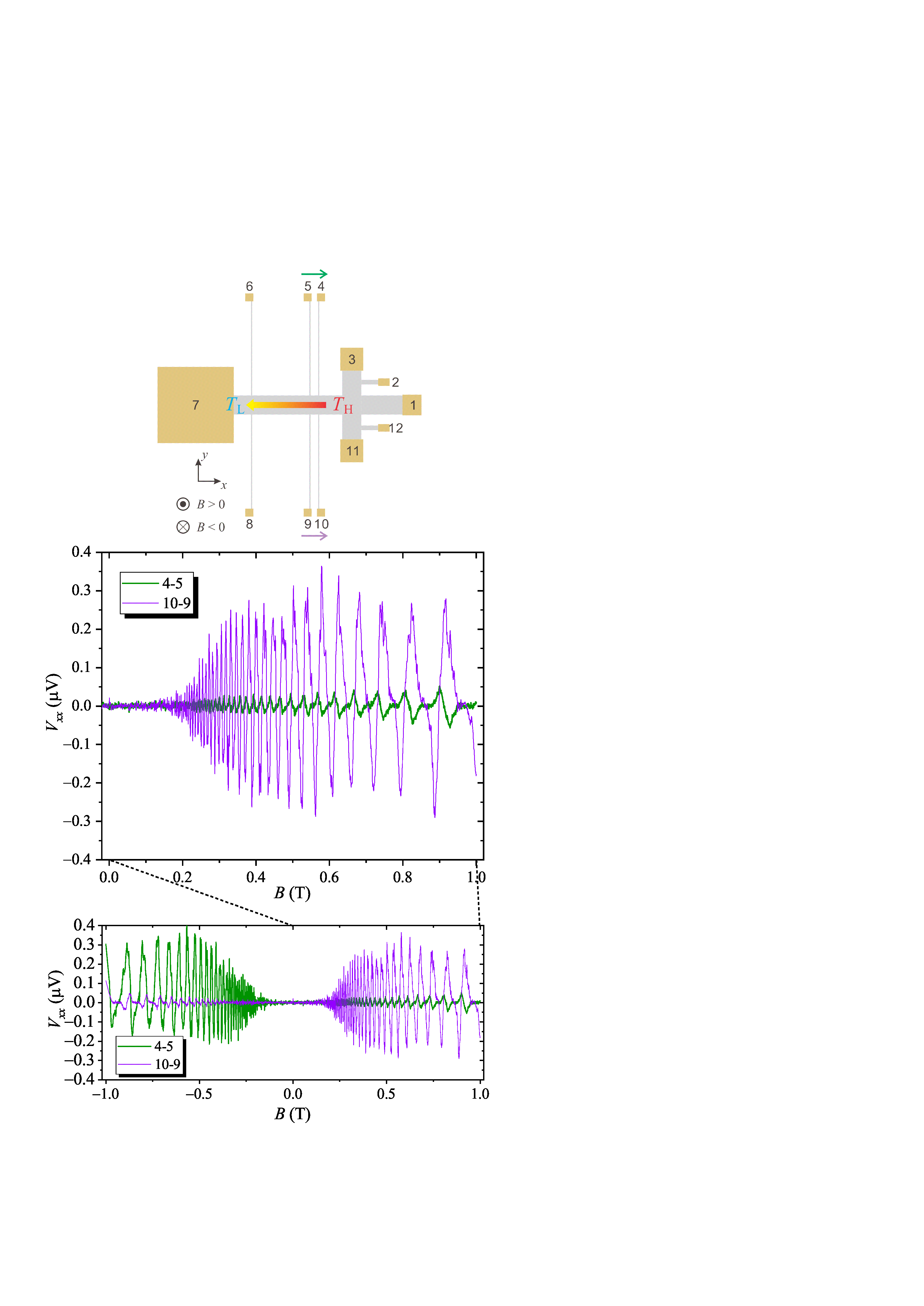} 
\caption{Magnetic-field dependence of the thermoelectric voltages $V_{xx}$ measured between the two pairs of probes, depicted by arrows in the top inset, located along the main Hall bar.  (The direction of the arrows indicates the sign of $V_{xx}$.) The bottom figure illustrates that the top (4-5) and the bottom (10-9) pairs interchange their roles by inverting the magnetic field.}
\label{MeasVxx}
\end{center}
\end{figure}

Figure \ref{MeasVxx} shows the longitudinal thermoelectric voltage $V_{xx}$ measured with pairs of voltage probes on the top (4-5) and the bottom (10-9) edges. Again, the quantum oscillations are observed above $\sim$0.2 T\@. Interestingly, the lineshape here also exhibits the basic traits of the off-diagonal thermoelectric voltage mentioned above, instead of showing the behaviour expected for the diagonal (Seebeck) thermoelectric voltage, characterized by a dip without sign reversal between two adjacent plateaus $V_{xx} \sim 0$ at integer fillings \cite{Jonson84} (see $S_{xx}$ in figure S2). The variation of the oscillation amplitude with $B$ is similar to that in $V_{yx}$. At $B > 0$, the top voltage-probe pair shows much smaller oscillation amplitude with the sign reversed compared to the bottom pair. By inverting the magnetic field ($B < 0$), however, the top and the bottom pairs switch their roles: the top (bottom) pair gains (loses) the oscillation amplitude and behaves like the mirror image of the trace for the bottom (top) pair in $B > 0$. In the following sections, we will interpret the behaviours of $V_{yx}$ and $V_{xx}$ described above in terms of the calculated spatial distribution of the electron temperature. 

\section{Calculation of the electron-temperature maps \label{calcT}}

In a GaAs/AlGaAs 2DES, the electron-phonon interaction is weak at low temperatures relevant to the present study,  as we will see in more detail below. We therefore start by neglecting the electron-phonon interaction. This allows us to obtain analytic solutions for the spatial distribution of the electron temperature $T$ for a one-dimensional (1D) model and for a rectangular sample, as will be briefly delineated in sections \ref{calcT1D} and \ref{calcTRect}, respectively. Numerical calculations resuming the electron-phonon interaction, performed on the Hall bar geometry used in the present study, will be presented in section \ref{calcTFEM}.

Neglecting the electron-phonon interaction, the continuity equation is written as,
\begin{equation}
\boldsymbol{\nabla}  \cdot {\bf{j}}_\mathrm{Q} = 0, \label{conteqwo}
\end{equation}
with ${\bf{j}}_\mathrm{Q} = - \hat{\kappa} \boldsymbol{\nabla} T$ the thermal flux density under the temperature gradient $\boldsymbol{\nabla} T$ and the  thermal conductivity tensor $\hat{\kappa}$. (We use ``hat'' to denote a tensor throughout the paper.) Applying the Wiedemann-Franz law (neglecting the thermal flux carried by phonons), $\hat{\kappa} =  L_0 T \hat{\sigma}$, with $L_0 = \pi ^2 k_\mathrm{B}^2 / (3 e^2) = 2.44 \times 10^{ - 8}$ W$\Omega$/K$^2$ the Lorenz number, $k_\mathrm{B}$ the Boltzmann constant and $\hat{\sigma}$ the conductivity tensor. Defining $\psi \equiv (T^2-{T_\mathrm{L}}^2)/2$, we have $\boldsymbol{\nabla} \psi \equiv T \boldsymbol{\nabla} T$ and
\begin{equation}
{\bf{j}}_\mathrm{Q} = -L_0 \hat{\sigma} \boldsymbol{\nabla} \psi \label{jQ}.
\end{equation}
Noting that the conductivity is virtually independent of the temperature below $\sim$1 K (the temperature range where the mobility is limited by impurity scattering) in a high-mobility GaAs/AlGaAs 2DES \cite{Davies98B}, and assuming a uniform 2DES, we can neglect the spatial derivative of $\hat{\sigma}$ even when placed in the spatially varying temperature. With further assumption that the 2DES is isotropic,  %$\sigma_{xx} = \sigma_{yy}$ and $\sigma_{yx} = - \sigma_{xy}$, 
we have $\boldsymbol{\nabla} \cdot (\hat{\sigma} \boldsymbol{\nabla} \psi) = \sigma_{xx} \boldsymbol{\nabla} ^2 \psi$. Thus, the equation to be solved becomes simply a Laplace's equation,
\begin{equation}
\boldsymbol{\nabla} ^2 \psi = 0. \label{Laplaceeq}
\end{equation}

\subsection{One-dimensional model \label{calcT1D}}
First, we consider a simplistic 1D model. In this model, we assume that the electron temperature $T$ is uniform across the  Hall bar ($y$ direction) and $T$ depends only on $x$ (the coordinate along the Hall bar, see figure \ref{HallBarFig} for the $x$-$y$ coordinate). We solve equation (\ref{Laplaceeq}) with the following boundary conditions: $T = T_\mathrm{L}$ at $x=0$ (the end of the Hall bar facing the heat sink) and $T = T_\mathrm{H}$ at $x = L$ (the other end, facing the heater). The solution $T(x)$ can readily be found:
\begin{equation}
T(x) = \left[ \left( {T_\mathrm{H}}^2 - {T_\mathrm{L}}^2 \right) \frac{x}{L} + T_\mathrm{L}^2 \right]^{1/2}. \label{Tx1D}
\end{equation}
%We will show below that this $T(x)$ yields a constant Nernst voltage $V_{yx}$ independent of $x$, which is in marked contradiction to our experimental results described in the previous section.
From equation (\ref{jQ}), we obtain the thermal flux density,
\begin{equation}
j_{\mathrm{Q}x} = -L_0 \sigma_{xx} \frac{{T_\mathrm{H}}^2 - {T_\mathrm{L}}^2}{2 L}. \label{jQ1D}
\end{equation}

\subsection{Two-dimensional analytic solution for a rectangular sample \label{calcTRect}}
\begin{figure}[tb]
\begin{center}
\includegraphics[width=8.3cm,clip]{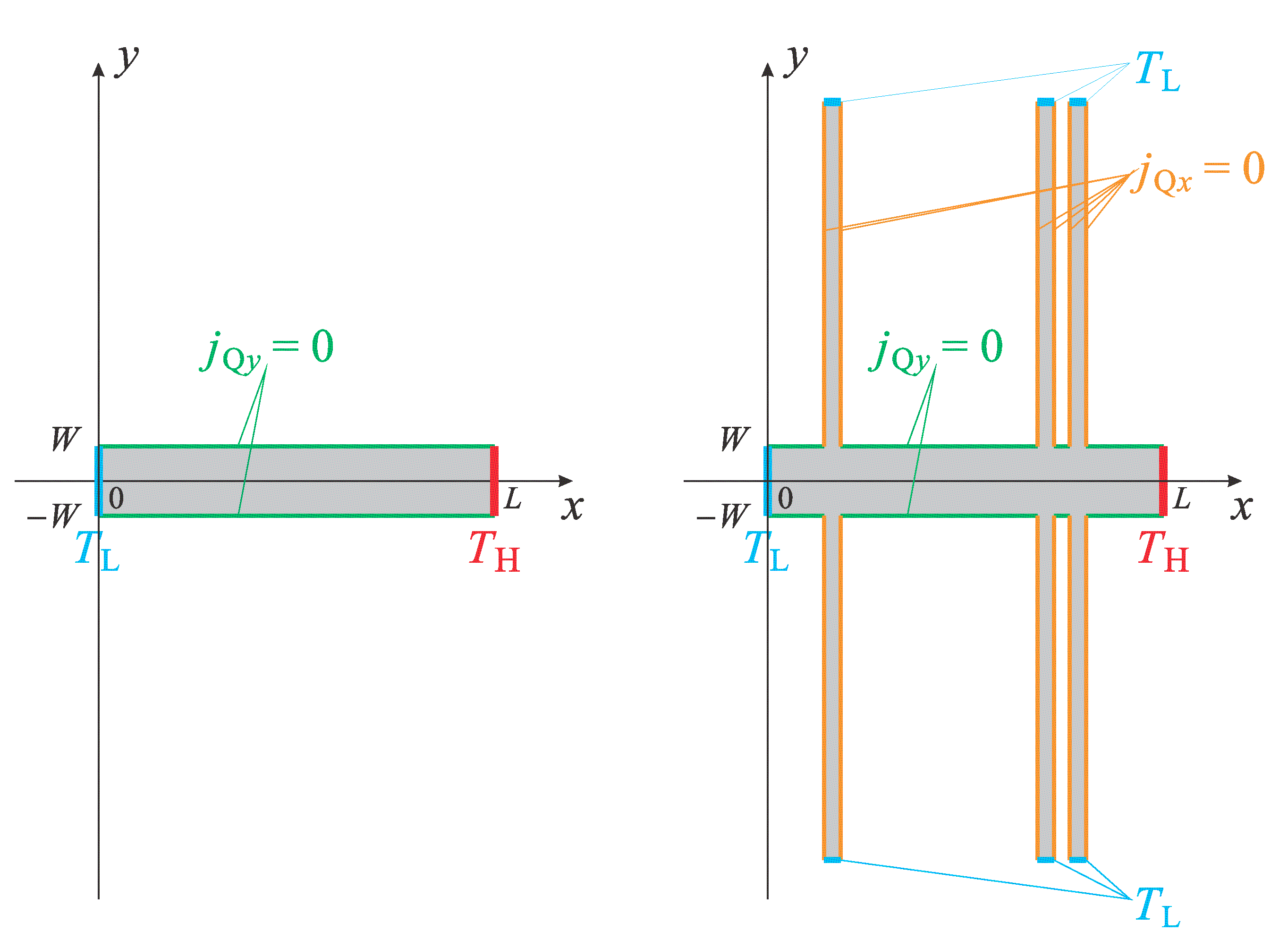} 
\caption{Boundary conditions for a rectangular sample (left) and the Hall bar geometry employed in the present study (right). The temperatures of the low-temperature and the high-temperature ends of the main Hall bar are fixed to $T_\mathrm{L}$ and $T_\mathrm{H}$, respectively. The temperature of the end of the voltage arms is also fixed to $T_\mathrm{L}$. The thermal flux is not permitted to cross the side edges. Refer to figure \ref{HallBarFig} for the dimensions of the sample. (Note that the right figure is not drawn to scale.)}
\label{boundary}
\end{center}
\end{figure}
Next, we take a look at the rectangular sample corresponding to the main part of the Hall bar with the voltage probes removed. The left panel of figure \ref{boundary} depicts the boundary conditions in this model.
% Defining the rectangular area in the $x$-$y$ plane as $0$ $\leq$ $x$ $\leq$ $L$ and $-W$ $\leq$ $y$ $\leq$ $W$, boundary conditions are given as follows: $T = T_\mathrm{L}$ at $x = 0$ and $-W$ $\leq$ $y$ $\leq$ $W$ (the end of the Hall bar facing the heat sink),  $T = T_\mathrm{H}$ at $x = L$ and  $-W$ $\leq$ $y$ $\leq$ $W$ (the other end, facing the heater), and $j_{\mathrm{Q}y} = 0$ at $0$ $\leq$ $x$ $\leq$ $L$ and either $y = -W$ or $y = W$ (top and bottom side edges of the Hall bar).
By further defining ${\bm{\tau}} \equiv -\boldsymbol{\nabla} \psi$, we can see that solving equation (\ref{Laplaceeq}) with these boundary conditions 
is mathematically equivalent to the process obtaining the electric field, reported by Rendell and Girvin \cite{Rendell81}, if we replace the electric field, the current density and the potential by ${\bm{\tau}}$, ${\bf{j}}_\mathrm{Q}/L_0$ and $\psi$, respectively. We can thus make use of the analytic solution presented in the paper \cite{Rendell81} (we follow the coordinate system used in \cite{Hirayama13}) to have
\numparts 
\begin{eqnarray}
\tau_x =  - \tau_0 e^\gamma \cos \vartheta \label{taux}, \\
\tau_y = \tau_0 e^\gamma \sin \vartheta, \label{tauy}
\end{eqnarray}
\endnumparts
and
\numparts
\begin{eqnarray}
j_{\mathrm{Q}x} =  - \tau_0 L_0 \frac{\sigma_{xx}}{\cos \delta} e^\gamma \cos \left( \vartheta  - \delta \right), \label{jQx} \\
j_{\mathrm{Q}y} = \tau_0 L_0 \frac{\sigma_{xx}}{\cos \delta} e^\gamma \sin \left( \vartheta  - \delta \right). \label{jQy}
\end{eqnarray}
\endnumparts
The parameters $\gamma$, $\vartheta$ and $\tau_0$ are given, using the Hall angle $\delta = \arctan (\sigma_{yx} / \sigma_{xx})$, as follows:  
\begin{equation}
\gamma  =  - 4 \delta \sum\limits_{n = 1}^\infty \frac{\sinh \left[ {\left( {2n - 1} \right)\pi y/L} \right] \cos \left[ {\left( {2n - 1} \right)\pi x/L} \right]}{{(2n - 1) \pi \cosh \left[ {\left( {2n - 1} \right)\pi \alpha/2} \right]}}, \label{gamma}
\end{equation}
\begin{equation}
\vartheta  = 4\delta \sum\limits_{n = 1}^\infty \frac{{\cosh \left[ {\left( {2n - 1} \right)\pi y/L} \right]\sin \left[ {\left( {2n - 1} \right)\pi x/L} \right]}}{{(2n - 1) \pi \cosh \left[ {\left( {2n - 1} \right)\pi \alpha/2}  \right]}} \label{vartheta}
\end{equation}
and
\begin{equation}
\tau _0 \left( \delta ,\alpha \right) = \frac{{T_\mathrm{H}}^2 - {T_\mathrm{L}}^2}{2 I\left( \delta ,\alpha \right)L}, \label{tau0}
\end{equation}
with
\begin{eqnarray}
I\left( {\delta ,\alpha } \right) \equiv \nonumber \\ 
 \int_0^1 \!\!\! {\cos \left\{ {4\delta \sum\limits_{n = 1}^\infty  {\frac{{\sin \left[ {\left( {2n - 1} \right)\pi \xi } \right]}}{{\left( {2n - 1} \right)\pi }}} {\mathop{\rm sech}\nolimits} \left[ {\left( {2n - 1} \right)\frac{{\alpha \pi }}{2}} \right]} \right\} d\xi }, \nonumber \\ \label{Iinteg}
\end{eqnarray}
where $\alpha \equiv 2W / L$ represents the aspect ratio. The boundary conditions can readily be confirmed by noting that $\vartheta = 0$ at $x = 0$, $L$ and $\vartheta = 4 \delta \Sigma_{n=1}^\infty \{ \sin[(2n-1)\pi x/L]/[(2n-1)\pi]  \} = \delta$ at $y = -W$, $W$\@.
The electron temperature can be obtained by integrating the ${\bm{\tau}}$ given above:
\begin{equation}
T(x,y) = \left[ {T_\mathrm{L}}^2 - 2 \int_0^x \tau_x(x^\prime,y) dx^\prime \right]^{1/2}. \label{TRect}
\end{equation}
%We will see below that the temperature calculated with this model captures the basic features of the spatial distribution of the temperature of a 2DES subjected to a perpendicular magnetic field. 
At $B = 0$, $\delta = \gamma = \vartheta =0$ and $I(0,\alpha) = 1$, and thus the temperature is given by equation (\ref{Tx1D}) regardless of the value of $y$.

We can find an approximate formula describing the total thermal flux
\begin{equation}
J_\mathrm{Q} = \int_{-W}^W j_{\mathrm{Q}x} dy \label{JQtot}
\end{equation}
flowing down the rectangle. Noting that $J_\mathrm{Q}$ is conserved along the $x$-direction, we evaluate equation (\ref{JQtot}) at $x = L/2$, where  $\gamma = 0$ and, for not too large $\alpha$, $\vartheta \simeq -4 \delta \Sigma_{n=1}^\infty \{ (-1)^n/[(2n-1)\pi] \} = \delta$. The thermal flux density $j_{\mathrm{Q}x} \simeq -\tau_0 L_0 \sigma_{xx} / \cos \delta$ becomes independent of $y$ and we have 
\begin{equation}
J_\mathrm{Q} = j_{\mathrm{Q}x} \cdot 2W \simeq - L_0 \sigma_{xx} \frac{\alpha}{I(\delta,\alpha) \cos \delta} \frac{{T_\mathrm{H}}^2 - {T_\mathrm{L}}^2}{2}. \label{JQRect}
\end{equation}
Since both $I(\delta,\alpha)$ and $\cos \delta$ decrease with increasing $\delta$, $J_\mathrm{Q}$ generally increases with $B$.
At $B = 0$, equation (\ref{JQRect}) can be simply rewritten as
\begin{equation}
J_\mathrm{Q} = j_{\mathrm{Q}x} \cdot 2W \simeq - L_0 \sigma_{xx} \alpha \frac{{T_\mathrm{H}}^2 - {T_\mathrm{L}}^2}{2}. \label{JQRectB0}
\end{equation}
and thus $j_{\mathrm{Q}x}$ coincide with that obtained by the 1D model, equation (\ref{jQ1D}). At a high magnetic field with $\delta$ approaching $\pi / 2$, on the other hand, we can show that $I(\delta,\alpha) \simeq \alpha$ for $\alpha < \sim 0.5$, resulting in
\begin{equation}
J_\mathrm{Q} \simeq - L_0 \frac{\sigma_{xx}}{\cos \delta} \frac{{T_\mathrm{H}}^2 - {T_\mathrm{L}}^2}{2} = L_0 \frac{\sin \delta}{\rho_{yx}} \frac{{T_\mathrm{H}}^2 - {T_\mathrm{L}}^2}{2}. \label{JQRectHighB}
\end{equation}
Note that $J_\mathrm{Q}$ does not depend on the aspect ratio $\alpha$ in equation (\ref{JQRectHighB}).

\subsection{Numerical calculation with electron-phonon interaction for a Hall bar sample \label{calcTFEM}}
\begin{figure*}[tb]
\begin{center}
\includegraphics[bb=15 190 810 690,width=17cm,clip]{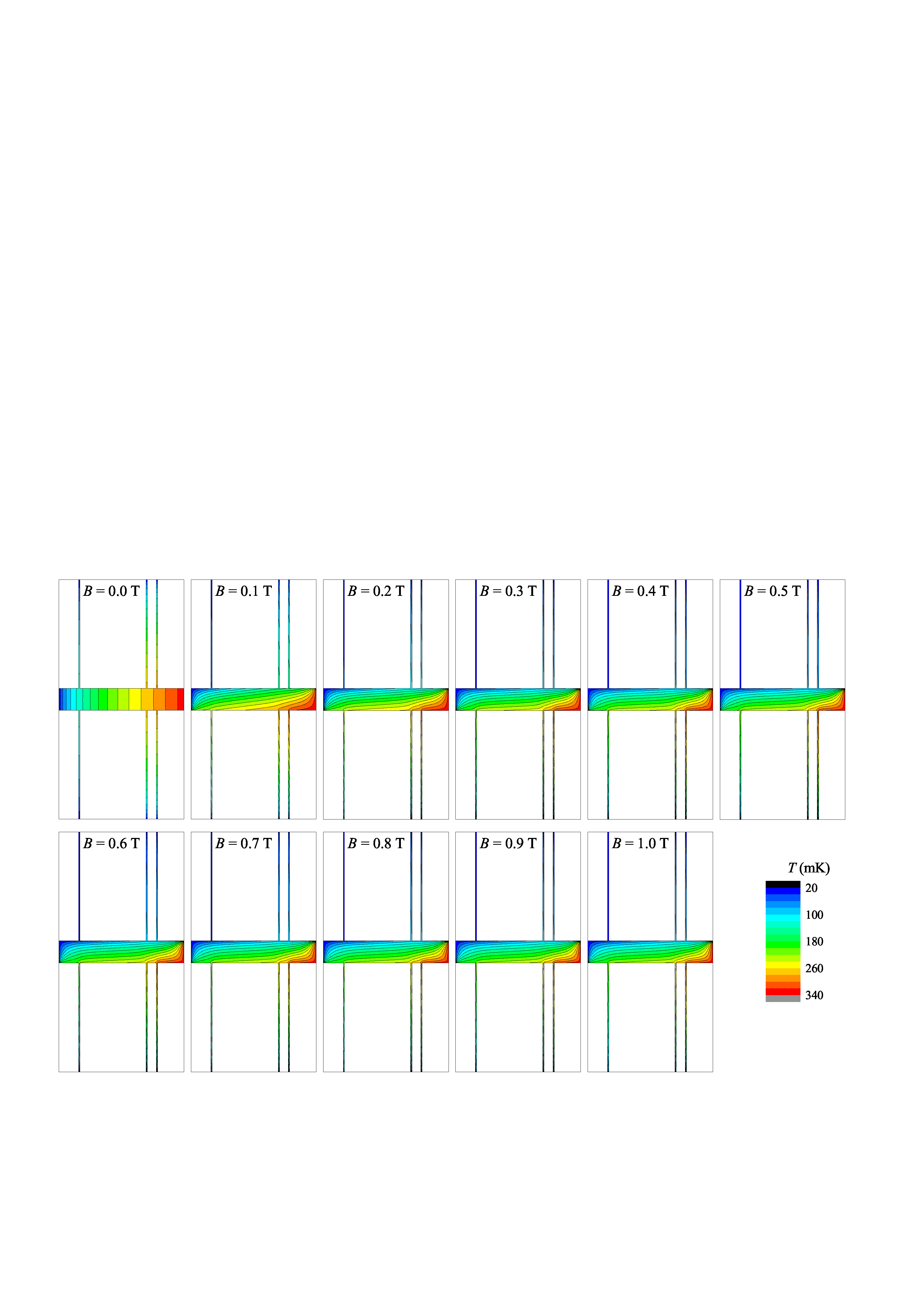} 
\caption{Spatial distribution of the electron temperature $T$ at various magnetic fields ranging from 0.0 to 1.0 T, calculated taking both the thermal diffusion into the voltage contacts and the electron-phonon interaction into consideration.}
\label{TAephcalc}
\end{center}
\end{figure*}
\begin{figure}[tb]
\begin{center}
\includegraphics[bb=30 0 570 600,width=8.3cm,clip]{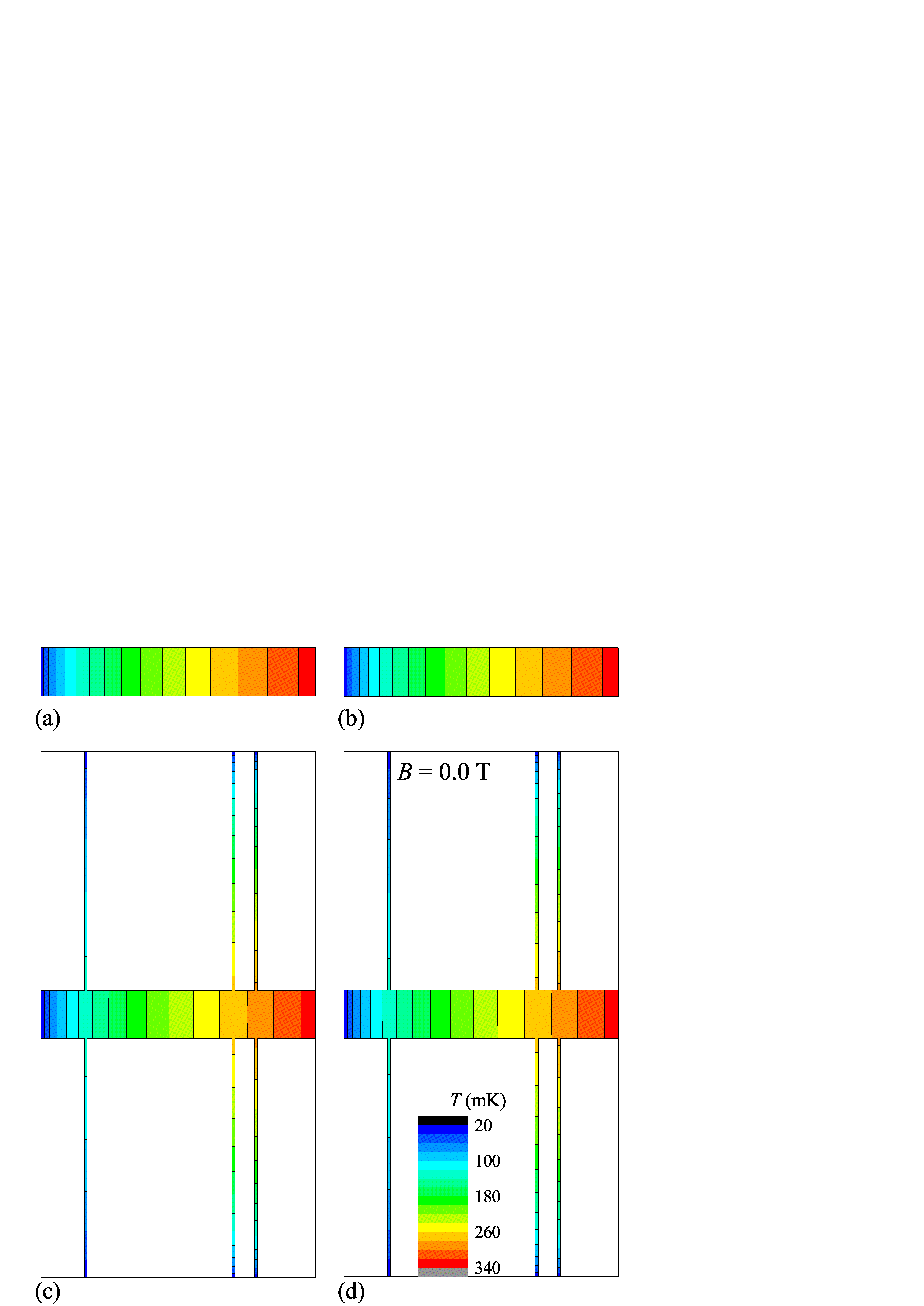} 
\caption{Spatial distribution of the electron temperature $T$ at $B = 0.0$ T\@. In (a) and (b), the voltage probes are neglected, and $T$ is calculated without (a) and with (b) the electron-phonon interaction ($T_\mathrm{(a)}$ and $T_\mathrm{(b)}$). In (c) and (d), the thermal diffusion into the voltage contacts is considered, and $T$ is calculated without (c) and with (d) the electron-phonon interaction ($T_\mathrm{(c)}$ and $T_\mathrm{(d)}$).}
\label{TB00calc}
\end{center}
\end{figure}
\begin{figure}[tb]
\begin{center}
\includegraphics[width=8.3cm,clip]{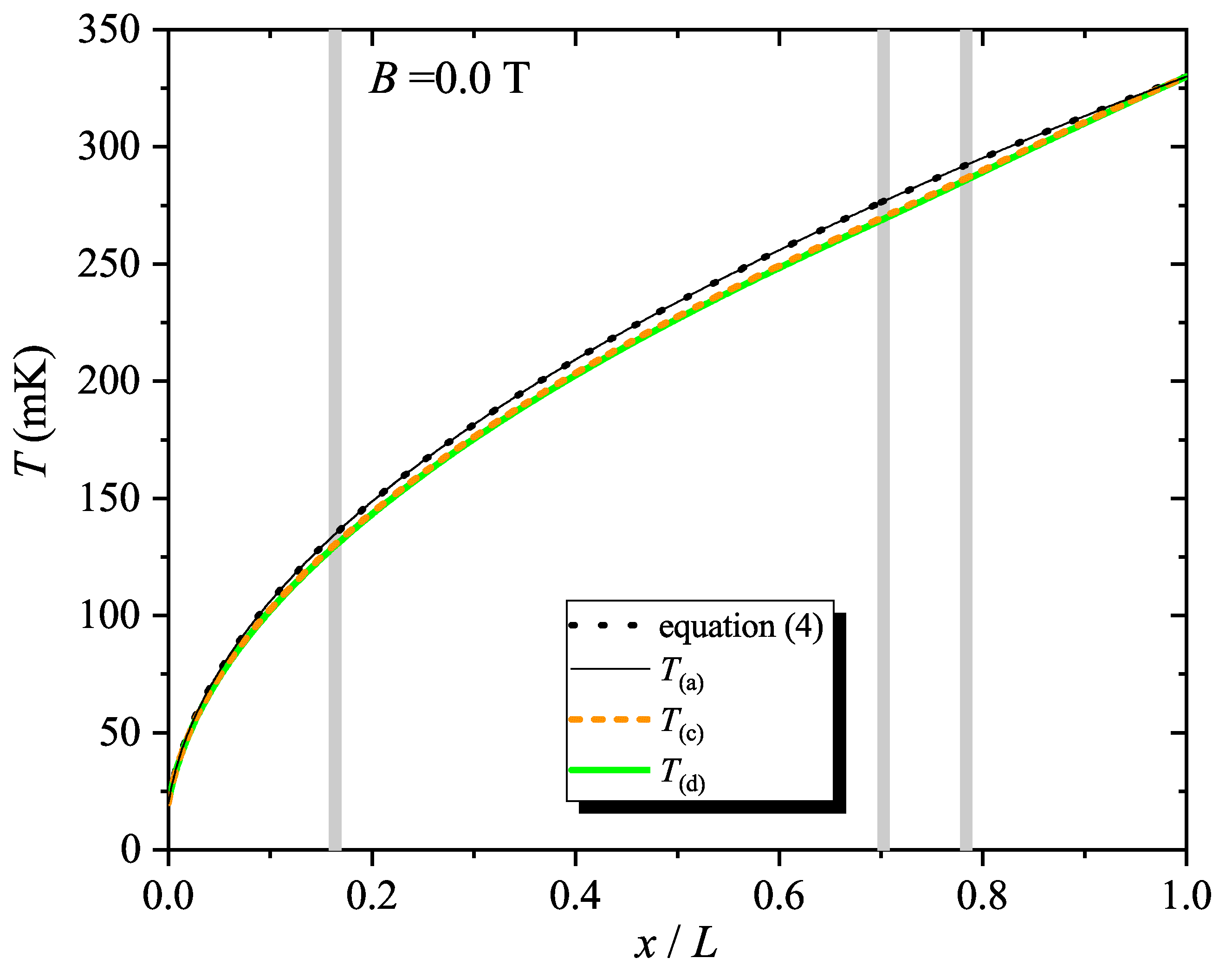}
\caption{Calculated electron temperature $T$ along the center line ($y = 0$, $0 \leq x \leq L$) of the main Hall bar at $B = 0.0$ T\@. Black dotted and thin solid lines, overlapping each other, represent the 1D model (equation (\ref{Tx1D})) and the numerical FEM calculation neglecting the electron-phonon interaction and using the rectangular boundary condition ($T_\mathrm{(a)}$), respectively. Thick orange dashed and green solid lines show the numerical FEM calculations for the Hall bar geometry without and with the electron-phonon interaction ($T_\mathrm{(c)}$ and $T_\mathrm{(d)}$), respectively. Gray shades indicate the positions of the voltage probes.}
\label{CSB00}
\end{center}
\end{figure}

\begin{figure}[tb]
\begin{center}
\includegraphics[bb=30 0 570 600,width=8.3cm,clip]{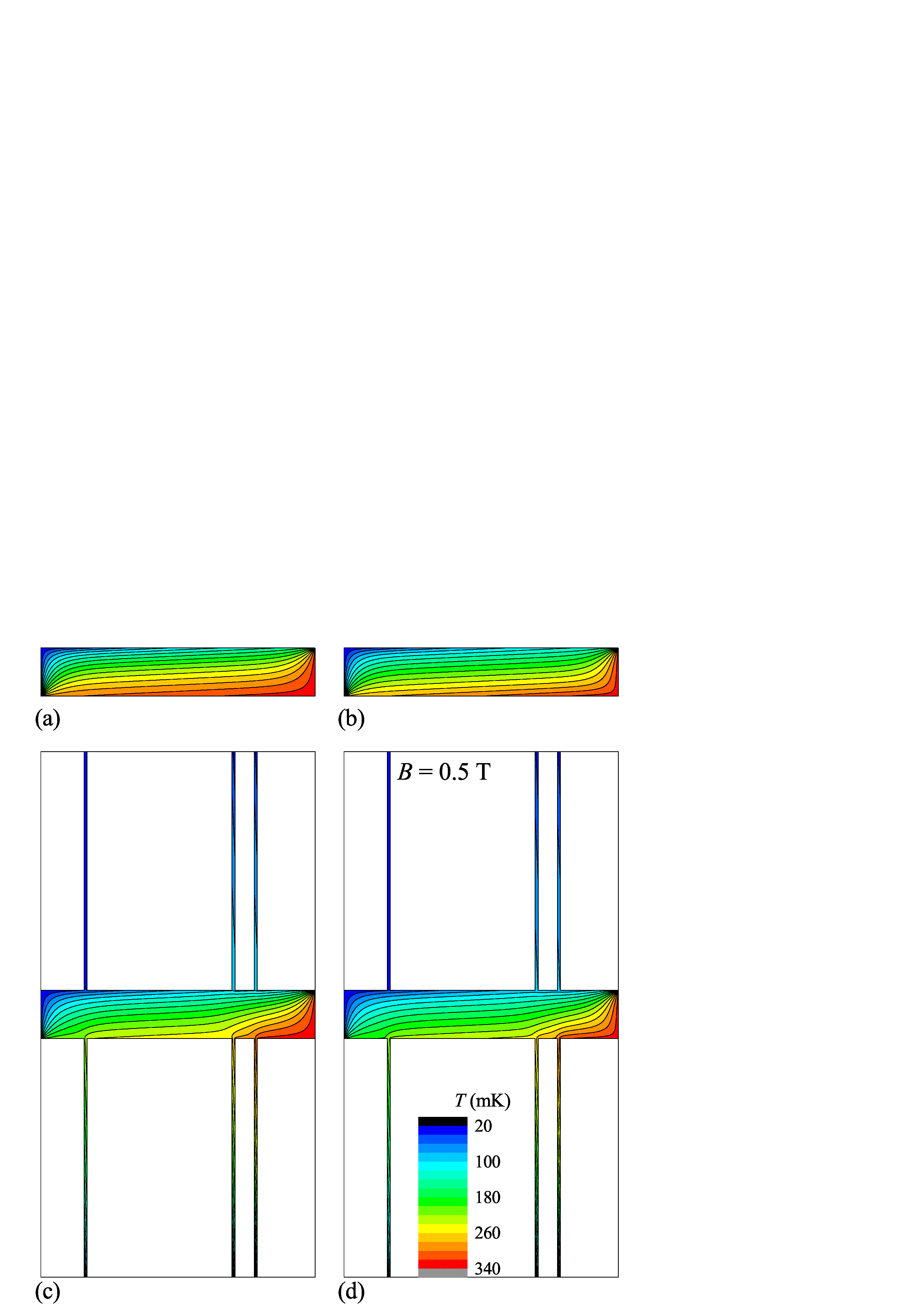} 
\caption{Spatial distribution of the electron temperature $T$ at $B = 0.5$ T\@. In (a) and (b), the voltage probes are neglected, and $T$ is calculated without (a) and with (b) the electron-phonon interaction ($T_\mathrm{(a)}$ and $T_\mathrm{(b)}$). In (c) and (d), the thermal diffusion into the voltage contacts is considered, and $T$ is calculated without (c) and with (d) the electron-phonon interaction ($T_\mathrm{(c)}$ and $T_\mathrm{(d)}$). }
\label{TB05calc}
\end{center}
\end{figure}
\begin{figure}[tb]
\begin{center}
\includegraphics[width=8.3cm,clip]{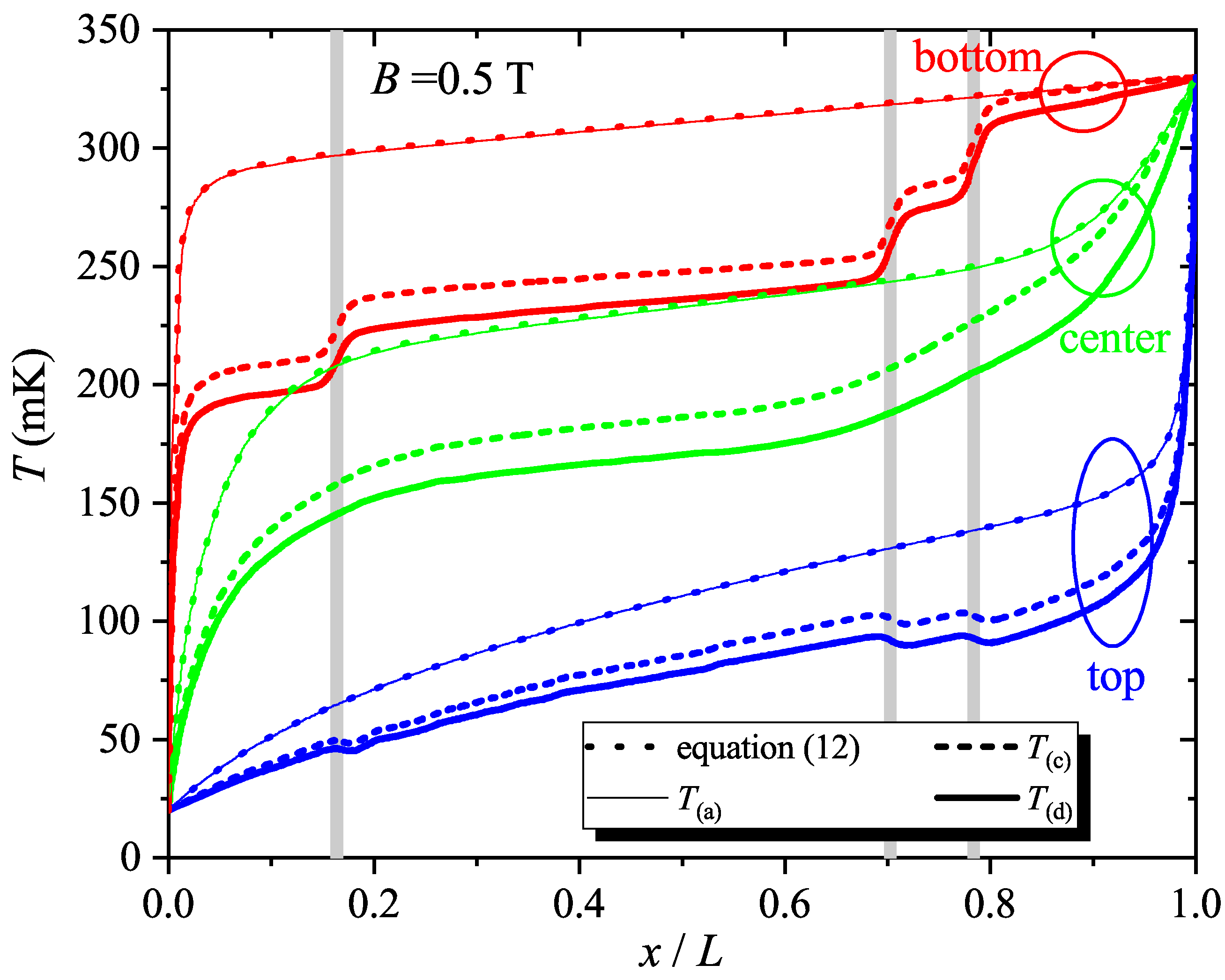}
\caption{Calculated electron temperature $T$ along the center line ($y = 0$, $0 \leq x \leq L$) and close to the bottom ($y = -W + w$, $0 \leq x \leq L$) and the top ($y = W - w$, $0 \leq x \leq L$) edges ($w = 1$ $\mu$m) of the main Hall bar at $B = 0.5$ T, plotted by green, red and blue lines respectively. Dotted and thin solid lines, overlapping each other, represent the analytic solution (equation (\ref{TRect})) and the numerical FEM calculation neglecting the electron-phonon interaction and using the rectangular boundary condition ($T_\mathrm{(a)}$), respectively. Thick dashed and solid lines show the numerical FEM calculations for the Hall bar geometry without and with the electron-phonon interaction ($T_\mathrm{(c)}$ and $T_\mathrm{(d)}$), respectively. Gray shades indicate the positions of the voltage probes.}
\label{CSB05}
\end{center}
\end{figure}

\begin{figure*}[tb]
\begin{center}
\includegraphics[bb=15 260 810 450,width=17cm,clip]{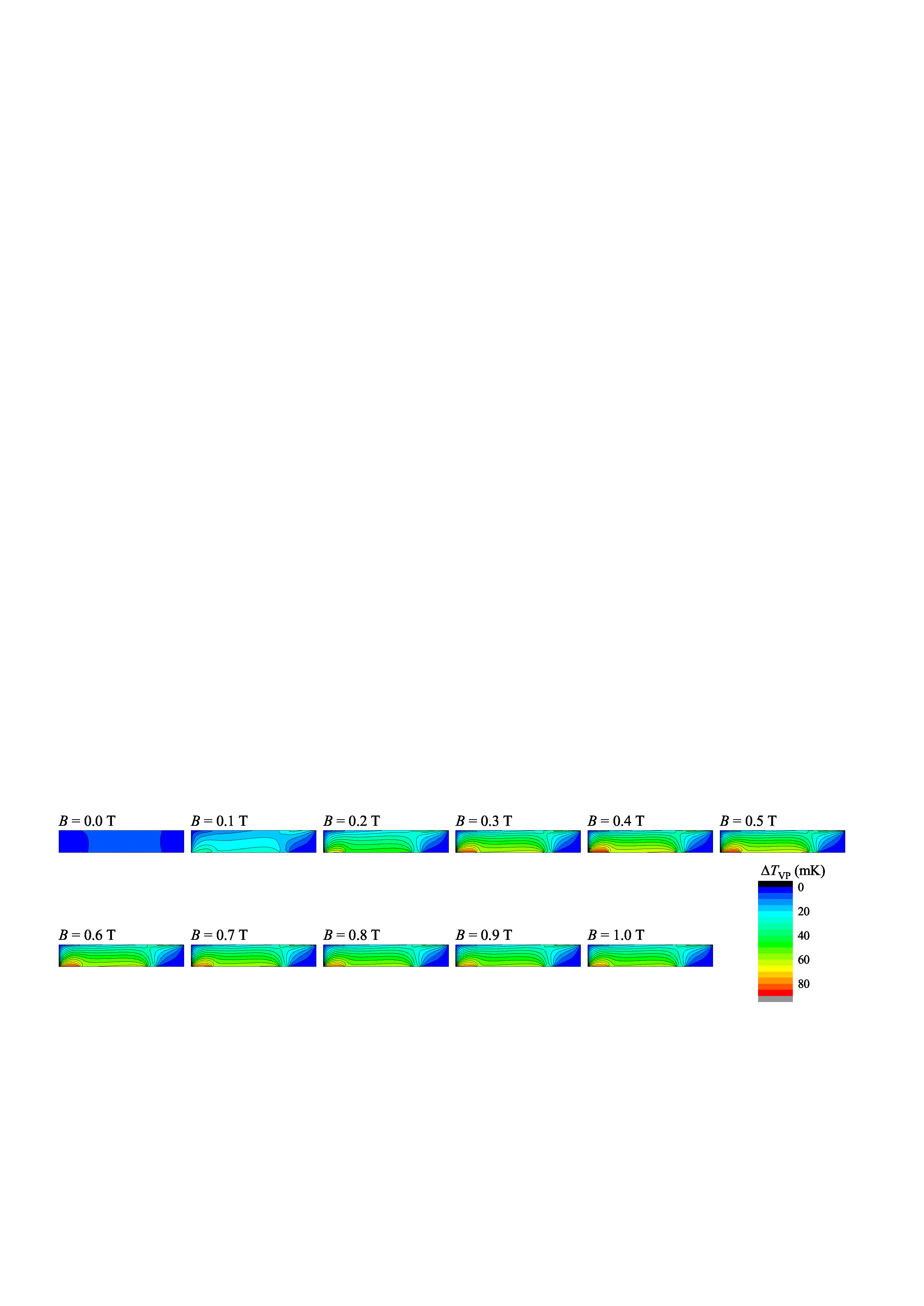} 
\caption{Spatial distribution of the temperature decrement due to the thermal diffusion into the voltage contacts, $\Delta T_\mathrm{VP} = T_\mathrm{(a)} - T_\mathrm{(c)}$, where $T_\mathrm{(a)}$ and $T_\mathrm{(c)}$ are the temperatures calculated without and with the thermal diffusion, respectively. The electron-phonon interaction is neglected in both  $T_\mathrm{(a)}$ and $T_\mathrm{(c)}$. }
\label{DTAcalc}
\end{center}
\end{figure*}

\begin{figure*}[tb]
\begin{center}
\includegraphics[bb=15 190 810 690,width=17cm,clip]{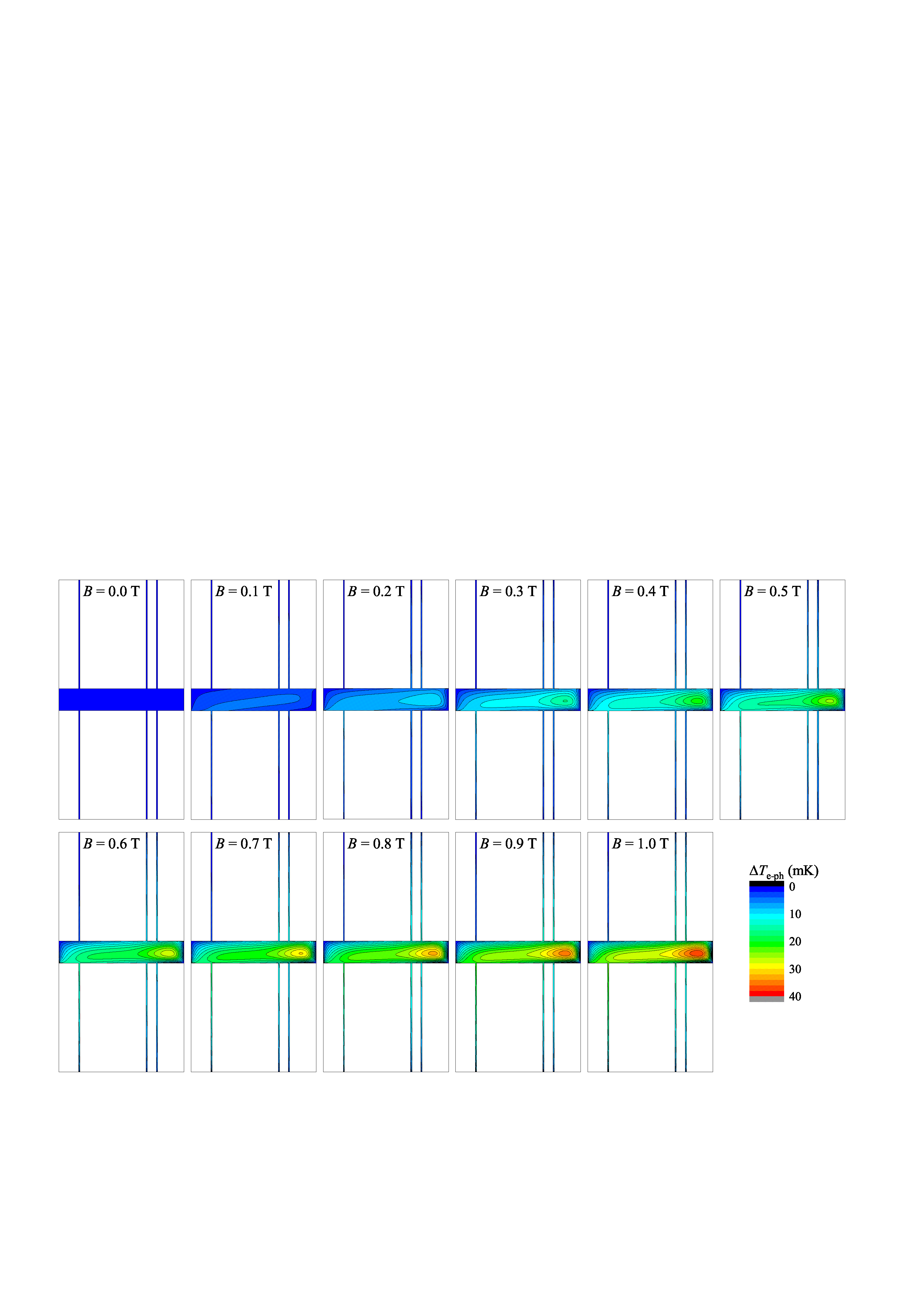} 
\caption{Spatial distribution of the temperature decrement due to the electron-phonon interaction, $\Delta T_\mathrm{e\mathchar`- ph} = T_\mathrm{(c)} - T_\mathrm{(d)}$, where $T_\mathrm{(c)}$ and $T_\mathrm{(d)}$ are the temperatures calculated without and with the electron-phonon interaction, respectively. The thermal diffusion into the voltage contacts is considered in both $T_\mathrm{(c)}$ and $T_\mathrm{(d)}$. }
\label{DTephcalc}
\end{center}
\end{figure*}
In order to examine the effect of the electron-phonon interaction as well as that of the voltage probes, we have to resort to numerical calculations. Incorporating the electron-phonon interaction, the continuity equation is altered from equation (\ref{conteqwo}) to  
\begin{equation}
\boldsymbol{\nabla}  \cdot {\bf{j}}_Q + P_\mathrm{e\mathchar`- ph}(T)  = 0, \label{conteqw}
\end{equation}
where
\begin{equation}
P_\mathrm{e\mathchar`- ph}(T) = P_\mathrm{def}(T) + P_\mathrm{pz}(T)
\end{equation}
represents the power (per area) transferred from the 2DES to the lattice via the electron-phonon interaction and is composed of two components: deformation-potential coupling $P_\mathrm{def} (T)$ and piezo-electric coupling $P_\mathrm{pz} (T)$ contributions \cite{Price82,KajiokaCO13}. They are approximately given by
\begin{equation}
P_\mathrm{def} (T) = \frac{P_\mathrm{D} \left( T^7 - {T_\mathrm{L}}^7 \right)}{{n_\mathrm{e}}^{1/2}} , \label{Pdef}
\end{equation}
and 
\begin{equation}
P_\mathrm{pz} (T) = \frac{P_\mathrm{P} \left( T^5 - {T_\mathrm{L}}^5 \right)}{{n_\mathrm{e}}^{1/2}}. \label{Ppz}
\end{equation}
Using the material parameters for GaAs, the coefficients are $P_\mathrm{D} \simeq 5.0 \times10^5$ W$\cdot$K$^{-7}$m$^{-3}$ and $P_\mathrm{P} \simeq 1.8 \times10^6$ W$\cdot$K$^{-5}$m$^{-3}$. Detailed derivation of equations (\ref{Pdef}) and (\ref{Ppz}) is presented in the Appendix. Replacing equation (\ref{jQ}) into equation (\ref{conteqw}), and again assuming the uniform and isotropic $\hat{\sigma}$, % and neglecting the spatial variation of $\hat{\sigma}$, 
we have
\begin{equation}
-L_0 \sigma_{xx} \boldsymbol{\nabla}^2 \psi + P_\mathrm{e\mathchar`- ph}(T) = 0. \label{epheq}
\end{equation}
We solve equation (\ref{epheq}) in the geometry of the main Hall bar used in the present study (enclosed in the light-green rectangle in figure \ref{HallBarFig}) using the boundary conditions illustrated in the right panel of figure \ref{boundary} with $T_\mathrm{L} = 20$ mK and  $T_\mathrm{H} = 330$ mK. %: $T = T_\mathrm{L}$ at the low-temperature end of the main Hall bar facing the heat-sink (contact 7) and at the end of the voltage arms facing the Ohmic contacts (contacts 4, 5, 6, 8, 9, 10), $T = T_\mathrm{H}$ at the high-temperature end of the main Hall bar, $j_{\mathrm{Q}y} = 0$ at the side edges of the main Hall bar and $j_{\mathrm{Q}x} = 0$  at the side edges of the voltage arms.

For the convenience of the numerical calculations, we rewrite equation (\ref{epheq}) in a dimensionless formula, using $L$ and $T_\mathrm{L}$ as units of the length and the temperature, respectively. We denote dimensionless quantities with the tilde: $\widetilde{T} \equiv T / T_\mathrm{L}$, $(\widetilde{x}, \widetilde{y}) \equiv (x/L, y/L)$, $\widetilde{\boldsymbol{\nabla}} = L \boldsymbol{\nabla}$, $\widetilde{\psi}  = ({\widetilde{T}}^2 - 1)/2$ and $\widetilde{\boldsymbol{\nabla}} \widetilde{\psi} = \widetilde{T} \widetilde{\boldsymbol{\nabla}} \widetilde{T} = (L/{T_\mathrm{L}}^2) \boldsymbol{\nabla} \psi$. By further introducing dimensionless coefficients,
\begin{equation}
\gamma _\mathrm{D} = \frac{P_\mathrm{D} L^2 {T_\mathrm{L}}^5}{L_0 \sigma_{xx} {n_e}^{1/2}} \label{gammaD}
\end{equation}
and
\begin{equation}
\gamma _\mathrm{P} = \frac{P_\mathrm{P} L^2 {T_\mathrm{L}}^3}{L_0 \sigma_{xx} {n_e}^{1/2}}, \label{gammaP}
\end{equation}
%equation (\ref{epheq}) is rewritten as
%\begin{equation}
%\widetilde{\boldsymbol{\nabla}}  \cdot \left( {\widetilde{\boldsymbol{\nabla}} \widetilde{\psi} } \right) - \gamma_\mathrm{D} \left( {{{\widetilde{T}}^7} - 1} \right) - \gamma_\mathrm{P} \left( {{{\widetilde{T}}^5} - 1} \right) = 0.
%\end{equation}
%and noting that we can set , or 
%\begin{equation}
%\widetilde{T} = \sqrt {2 \widetilde{\psi} + 1} \label{tildeT}
%\end{equation}
%without loss of generality by a suitable choice of the integral constant, 
equation (\ref{epheq}) is rewritten as
\begin{eqnarray}
\widetilde{\boldsymbol{\nabla}} ^2 \widetilde{\psi}  - \gamma_\mathrm{D} \left[ \left( {2 \widetilde{\psi}  + 1} \right)^{7/2} - 1 \right] - \gamma_\mathrm{P} \left[ \left( {2 \widetilde{\psi}  + 1} \right)^{5/2} -1 \right] \nonumber \\ \hspace{60mm} = 0. \label{psieq}
\end{eqnarray}
We numerically solve equation (\ref{psieq}) by the finite element method (FEM) using FreeFem++ \cite{Hecht12}. %under the boundary conditions mentioned above 
Since we are primarily interested in the amplitude of the quantum oscillations in the thermoelectric voltages, we used the upper envelop of the SdH oscillations ($\sigma_\mathrm{UE}$ presented in figure S3 in the Supplementary data) as the $\sigma_{xx}$ to be substituted into equations (\ref{gammaD}) and (\ref{gammaP}). 
%Semiclassical $\sigma_{xx} = \sigma_0 / (1 + \mu^2 B^2)$

The resulting maps of the electron temperature 
\begin{equation}
T = T_\mathrm{L} (2 \widetilde{\psi} + 1)^{1/2} \label{TFEM}
\end{equation}
are presented in figure \ref{TAephcalc} for the magnetic field $B =0.0$, 0.1, .... 1.0 T\@. At $B = 0.0$ T, the temperature exhibits monotonic decrease from the heater to the heat sink. Once a magnetic field is applied, however, the temperature distribution becomes heavily distorted. This is mainly attributable to the large Hall angle $\delta$ %in a high-mobility 2DES, 
%rapidly  approaching $\pi/2$ with a relatively low magnetic field 
(the plot of $\tan \delta$ vs.\ $B$ used in the present calculation is shown in figure S3 in the Supplementary data). The large $\delta$ along with the boundary condition that the thermal flux $\bf j_\mathrm{Q}$ should not cross the top and the bottom edges of the Hall bar require the temperature gradient $\boldsymbol{\nabla} T$ to have a large angle with the $x$ axis in the main Hall bar (see equation (\ref{jQ})). The spatial distribution of the temperature does not vary noticeably above $\sim$0.3 T, where $\delta$ becomes practically saturated.

To elucidate how the voltage probes and the electron-phonon interaction have affected the temperature maps shown in figure \ref{TAephcalc}, we repeated the numerical FEM calculations eliminating these effects. We calculated the electron temperature $T$ in the following cases: (a) a rectangular sample without the voltage probes and without the electron-phonon interaction, (b) a rectangular sample without the voltage probes but with the electron-phonon interaction, (c) a Hall bar sample with the voltage probes but without electron-phonon interaction, in addition to (d) a Hall bar sample with the voltage probes and with the electron-phonon interaction (shown in figure \ref{TAephcalc}). The results of the calculations $T_\mathrm{(a)}$, $T_\mathrm{(b)}$, $T_\mathrm{(c)}$ and $T_\mathrm{(d)}$ at $B = 0.0$ T and at $B = 0.5$ T are presented in figures \ref{TB00calc} %(a)--(d) and figure 
and \ref{TB05calc}, %(a)--(d), 
respectively, and some of their selected cross sections are plotted in figures \ref{CSB00} and \ref{CSB05}, respectively. In order to explicitly visualize the effect of the voltage probes and the electron-phonon interaction, we plot the temperature decrement owing to these effects, $\Delta T_\mathrm{VP} = T_\mathrm{(a)} - T_\mathrm{(c)}$ and $\Delta T_\mathrm{e\mathchar`- ph} = T_\mathrm{(c)} - T_\mathrm{(d)}$, in figures \ref{DTAcalc} and \ref{DTephcalc}, respectively.

At $B = 0.0$ T, $T_\mathrm{(a)}$ reproduces $T(x)$ in the 1D model (equation (\ref{Tx1D})) independent of $y$ (see figures \ref{TB00calc} (a) and \ref{CSB00}). The effects of the voltage probes and the electron-phonon interaction are small and are not readily discernible in figure \ref{TB00calc}. Isothermal lines are seen to bend, albeit very faintly, near the voltage probes in $T_\mathrm{(c)}$ and $T_\mathrm{(d)}$. The cross section at $y = 0$ shown in figure \ref{CSB00} and the temperature decrement plotted in figures \ref{DTAcalc} and \ref{DTephcalc} reveal that voltage probes slightly reduces the temperature ($T_\mathrm{(c)}$), while further inclusion of the electron-phonon interaction barely alters the temperature ($T_\mathrm{(d)}$). The very small disturbance of the temperature distribution by the voltage probes attests to the success in the design of the thin and long arms. The almost negligible effect of the electron-phonon interaction is as expected at low temperatures from the $T$ dependence $P_\mathrm{def} \sim T^7$ and $P_\mathrm{pz} \sim T^5$. 

A magnetic field considerably enhances the disturbances, as clearly visualized in figure \ref{CSB05} for an example at $B = 0.5$ T\@. First, we can confirm that $T_\mathrm{(a)}$ replicates the analytic solution $T(x,y)$ for a rectangular sample, equation (\ref{TRect}). A substantial drop of the temperature is induced by the voltage contacts ($T_\mathrm{(c)}$), with the decrement $\Delta T_\mathrm{VP}$ becoming more pronounced as we go closer to the bottom (higher-temperature) edge (see also figure \ref{DTAcalc}). Slight reduction $\Delta T_\mathrm{e\mathchar`- ph}$ is further made by the electron-phonon interaction ($T_\mathrm{(d)}$). Figure \ref{DTephcalc} illustrates that $\Delta T_\mathrm{e\mathchar`- ph}$ is amplified by the increase of the magnetic field.

By applying the analytic rectangular model described in section \ref{calcTRect} to the arms of the voltage probe, we can qualitatively understand why the voltage probes cool down the main bar more effectively in the magnetic field. Without the magnetic field, the thermal flux through the arm $J_\mathrm{Q}$ can be made arbitrarily small by reducing the aspect ratio $\alpha$, as indicated by equation (\ref{JQRectB0}). As mentioned earlier, $J_\mathrm{Q}$ increases with $B$, and the strategy of reducing $\alpha$ ceases to be effective at the magnetic field with $\delta \sim \pi/2$, where $J_\mathrm{Q}$ becomes insensitive to $\alpha$ as implied in equation (\ref{JQRectHighB}).
The increase of $\Delta T_\mathrm{e\mathchar`- ph}$ with the increase of the magnetic field is attributable to the decrease in $\sigma_{xx} = \sigma_\mathrm{UE}$, which enhances the coefficients $\gamma_\mathrm{D}$ and $\gamma_\mathrm{P}$ in equations (\ref{gammaD}) and (\ref{gammaP}). Comparison of figure \ref{DTAcalc} and figure \ref{DTephcalc} reveals that the effect of the voltage probes generally outweighs that of the electron-phonon interaction in reducing the electron temperature in our experimental conditions (note the difference in the colour code scales in the two figures).

\section{Simulating thermoelectric voltages from the calculated spatial distribution of the electron temperature \label{thermvol}}
\begin{figure}[tb]
\begin{center}
\includegraphics[width=8.3cm,clip]{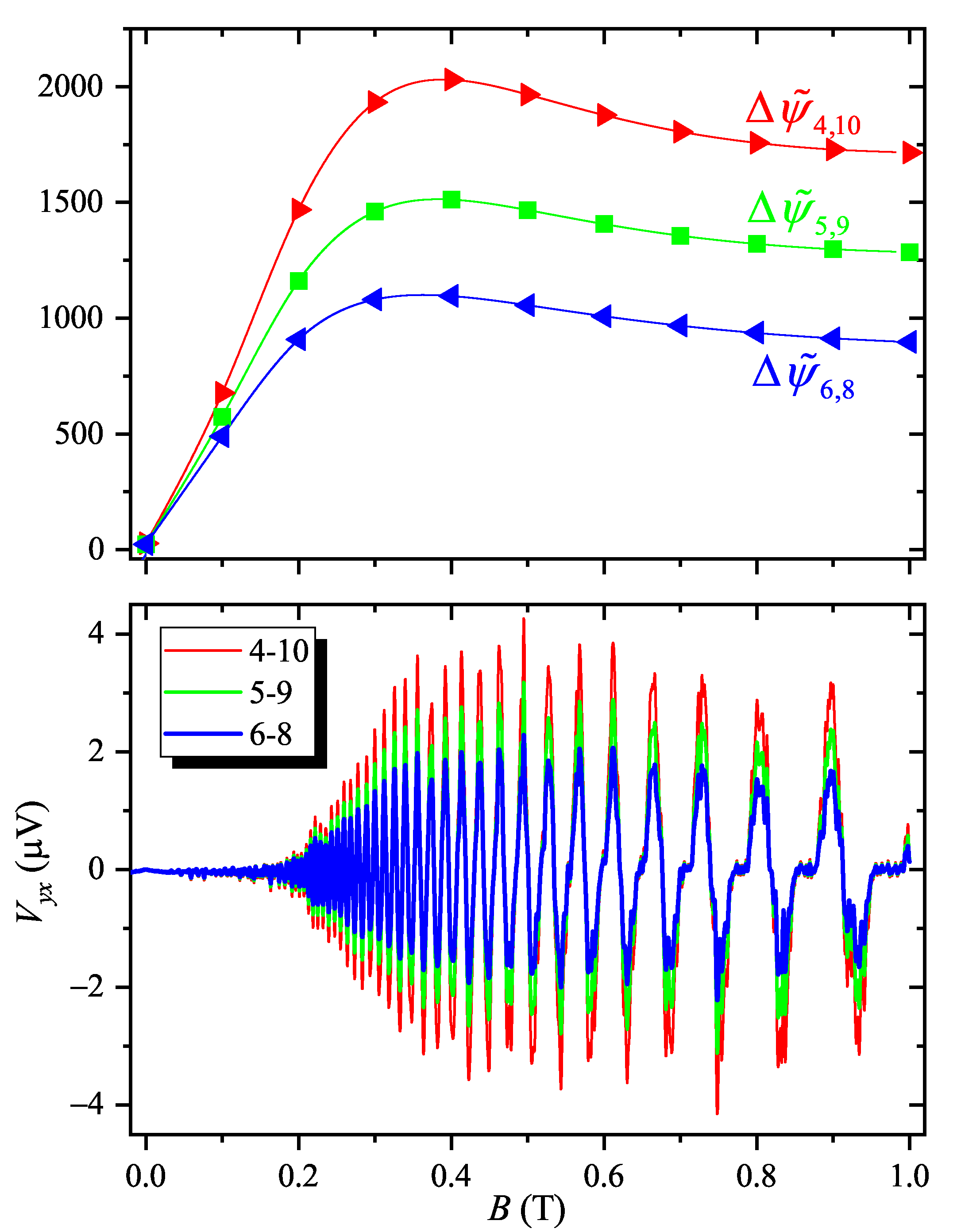} 
\caption{Top: $\Delta \widetilde{\psi}_{i,j}$ given by equation (\ref{Dpsiyx}) at $B =0.0$, 0.1, ... 1.0 T calculated using the spatial distribution of $T$ %the electron temperature 
presented in figure \ref{TAephcalc}. Red right-pointing triangles, green squares and blue left-pointing triangles represent $(i,j) = (4,10)$, $(5,9)$ and $(6,8)$, respectively, and lines are the interpolating spline curves. Bottom: Magnetic-field dependence of the thermoelectric voltages between contacts $i$ and $j$, $V_{yx} = \phi_{i,j}$, calculated using equation (\ref{phiijc}) with the (interpolated) $\Delta \widetilde{\psi}_{i,j}$ shown in the top panel and $s_{yx} = S_{yx} / T$ deduced from experimentally obtained $\rho_{xx}(B)$ and $\rho_{yx}(B)$.}
\label{SimVyx}
\end{center}
\end{figure}

\begin{figure}[tb]
\begin{center}
\includegraphics[width=8.3cm,clip]{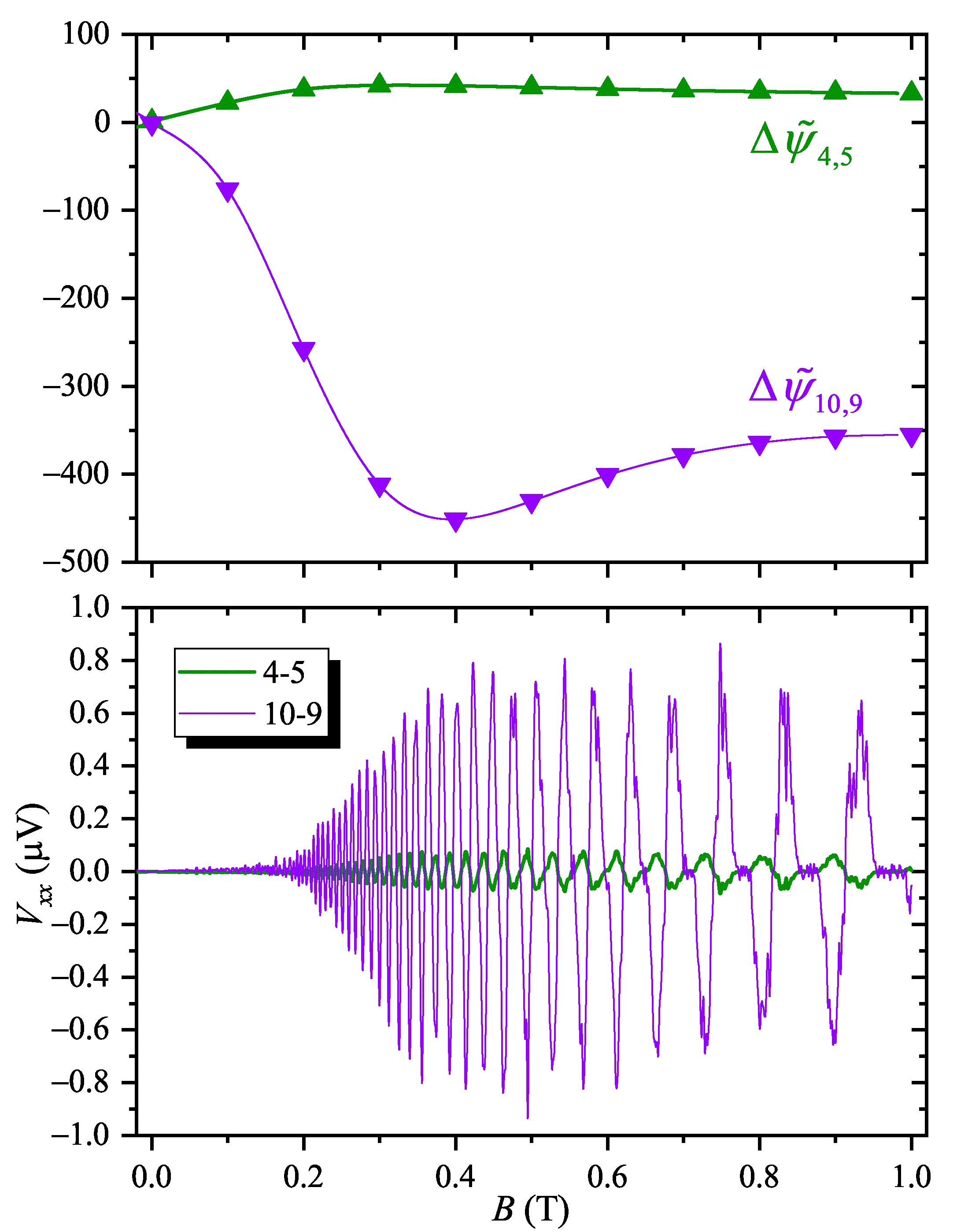} 
\caption{Top: $\Delta \widetilde{\psi}_{i,j}$ given by equation (\ref{Dpsixx}) at $B =0.0$, 0.1, ... 1.0 T calculated using the spatial distribution of $T$ %the electron temperature 
presented in figure \ref{TAephcalc}. Dark-green upright triangles and purple downward triangles represent $(i,j) = (4,5)$ and $(10,9)$, respectively, and lines are the interpolating spline curves. Bottom: Magnetic-field dependence of the thermoelectric voltages between contacts $i$ and $j$, $V_{xx} = \phi_{i,j}$, calculated using equation (\ref{phiijc}) with the (interpolated) $\Delta \widetilde{\psi}_{i,j}$ shown in the top panel and $s_{yx} = S_{yx} / T$ deduced from experimentally obtained $\rho_{xx}(B)$ and $\rho_{yx}(B)$.}
\label{SimVxx}
\end{center}
\end{figure}

In this section, we evaluate the thermoelectric voltage resulting from the gradient $\boldsymbol{\nabla} T$ of the temperature calculated in section \ref{calcT}. We compare the calculated thermoelectric voltages with those measured experimentally, presented in section \ref{meas}, focusing on the amplitude of the quantum oscillations.

Local electric field ${\bf E}$ induced by $\boldsymbol{\nabla} T$ is given by
\begin{equation}
{\bf E} = \hat{S} \boldsymbol{\nabla} T, \label{ESnablaT}
\end{equation}
where $\hat{S}$ represents the thermopower tensor. Since $k_\mathrm{B} T \ll E_\mathrm{F}$ in the temperature range studied in the present paper, we can make use of the generalized Mott's formula (applicable to the 2DES subjected to a perpendicular magnetic field)  \cite{Jonson84},
\begin{equation}
\hat{S} = -e L_0 T \hat{\sigma}^{-1} \left. \frac{\partial \hat{\sigma}}{\partial E} \right|_{E = E_\mathrm{F}} = T \hat{s} \label{Mott}.
\end{equation}
Here we introduced $\hat{s} \equiv \hat{S} / T$  for convenience, whose temperature dependence can be neglected under the assumption of the temperature-independent $\hat{\sigma}$. With this, we have
\begin{equation}
{\bf E} = \hat{s} \boldsymbol{\nabla} \psi. \label{Esnablapsi}
\end{equation}
We can readily see that the temperature distribution in the 1D model, equation (\ref{Tx1D}), results in $\boldsymbol{\nabla} \psi = \partial \psi / \partial x =$ constant. Therefore ${\bf E}$, and accordingly $V_{yx}$, does not vary with $x$ in this model. The diminishing oscillation amplitude with decreasing $x$ exhibited by the quantum oscillations of $V_{yx}$ in figure \ref{MeasVyx} is thus clearly irreconcilable with the purely 1D model.

Using the resistivity tensor $\hat{\rho} = \hat{\sigma}^{-1}$, equation (\ref{Mott}) can be rewritten as,
\numparts 
\begin{eqnarray} 
s_{xx} = s_{yy} = \frac{S_{xx}}{T} = e L_0 \left. \frac{d}{dE} \ln \sqrt{{\rho _{xx}}^2 + {\rho _{yx}}^2} \right|_{E = E_F},  \label{Srhoxx} \\
s_{yx} = -s_{xy} = \frac{S_{yx}}{T} = e L_0 \left. \frac{d}{dE} \arctan \frac{\rho_{yx}}{\rho_{xx}} \right|_{E = E_F}, \label{Srhoyx}
\end{eqnarray}
\endnumparts
where we made use of the relations $\rho_{xx} = \rho_{yy}$ and $\rho_{xy} = -\rho_{yx}$ fulfilled in an isotropic 2DES\@. %The relations result in $s_{xx} = s_{yy}$ and $s_{xy} = -s_{yx}$.
Noting that the quantum oscillations are resulting from the location of the Fermi energy with respect to the Landau levels, we can replace the energy derivative by the derivative with respect to the magnetic field,  
\[ \frac{d}{dE} \to  - \frac{B}{E_\mathrm{F}} \frac{d}{dB}, \]
for the assessment of oscillation amplitude. This allows us to deduce the components of $\hat{S}$ from experimentally obtained $\rho_{xx}(B)$ and $\rho_{yx}(B)$ (shown in figure S1 in the Supplementary data), using equations (\ref{Srhoxx}) and (\ref{Srhoyx}). The resulting $S_{xx}$ and $S_{yx}$ are also presented in the Supplementary data (figure S2). 

\begin{figure}[tb]
\begin{center}
\includegraphics[width=8.3cm,clip]{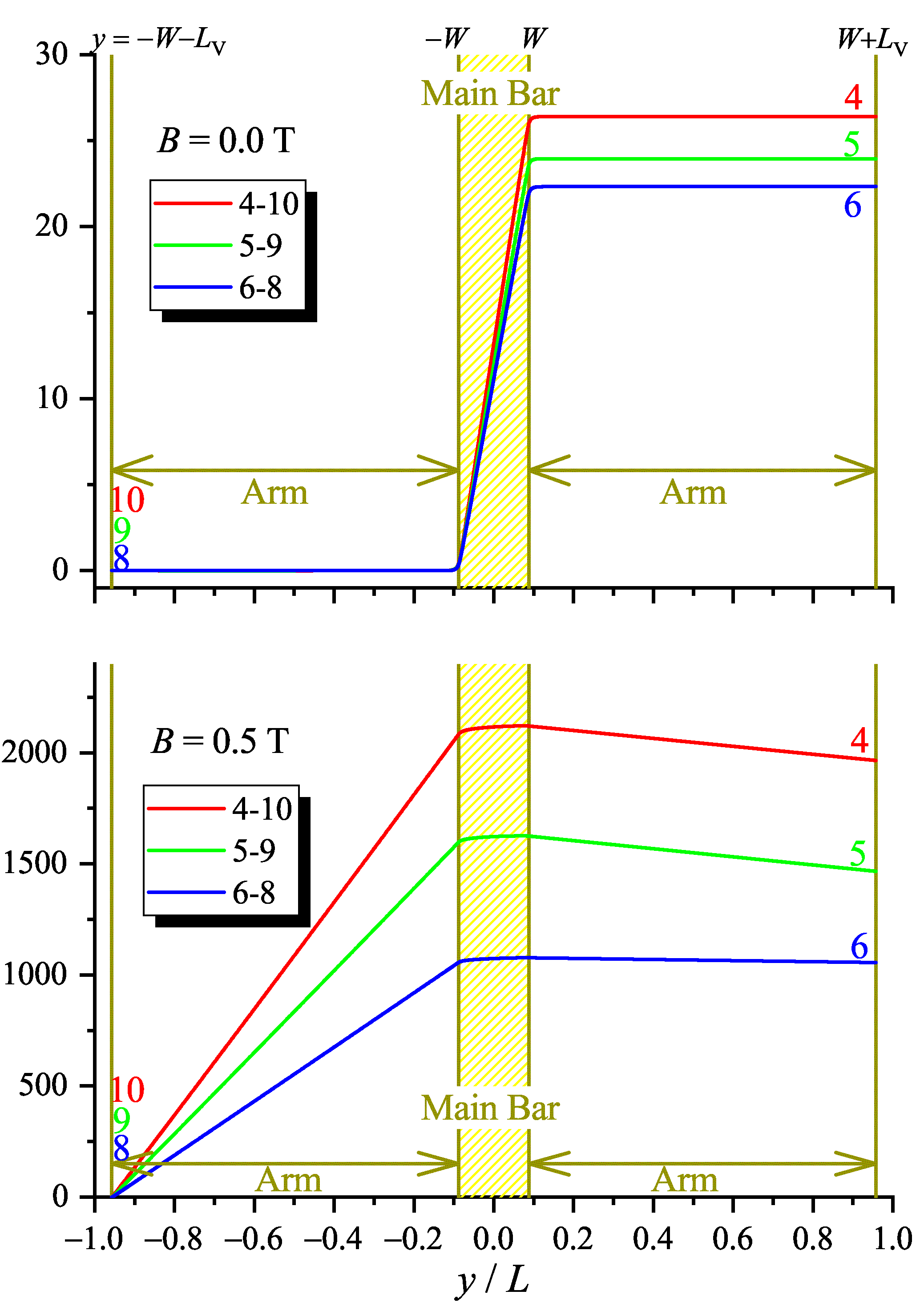} 
\caption{$\Delta \widetilde{\psi}_{yx}(\widetilde{y})$ given by equation (\ref{psiyx}) for contact pairs 6-8, 5-9 and 4-10 at $B =0.0$ T (top) and $B = 0.5$ T (bottom). }
\label{Integyx}
\end{center}
\end{figure}

\begin{figure}[tb]
\begin{center}
\includegraphics[width=8.3cm,clip]{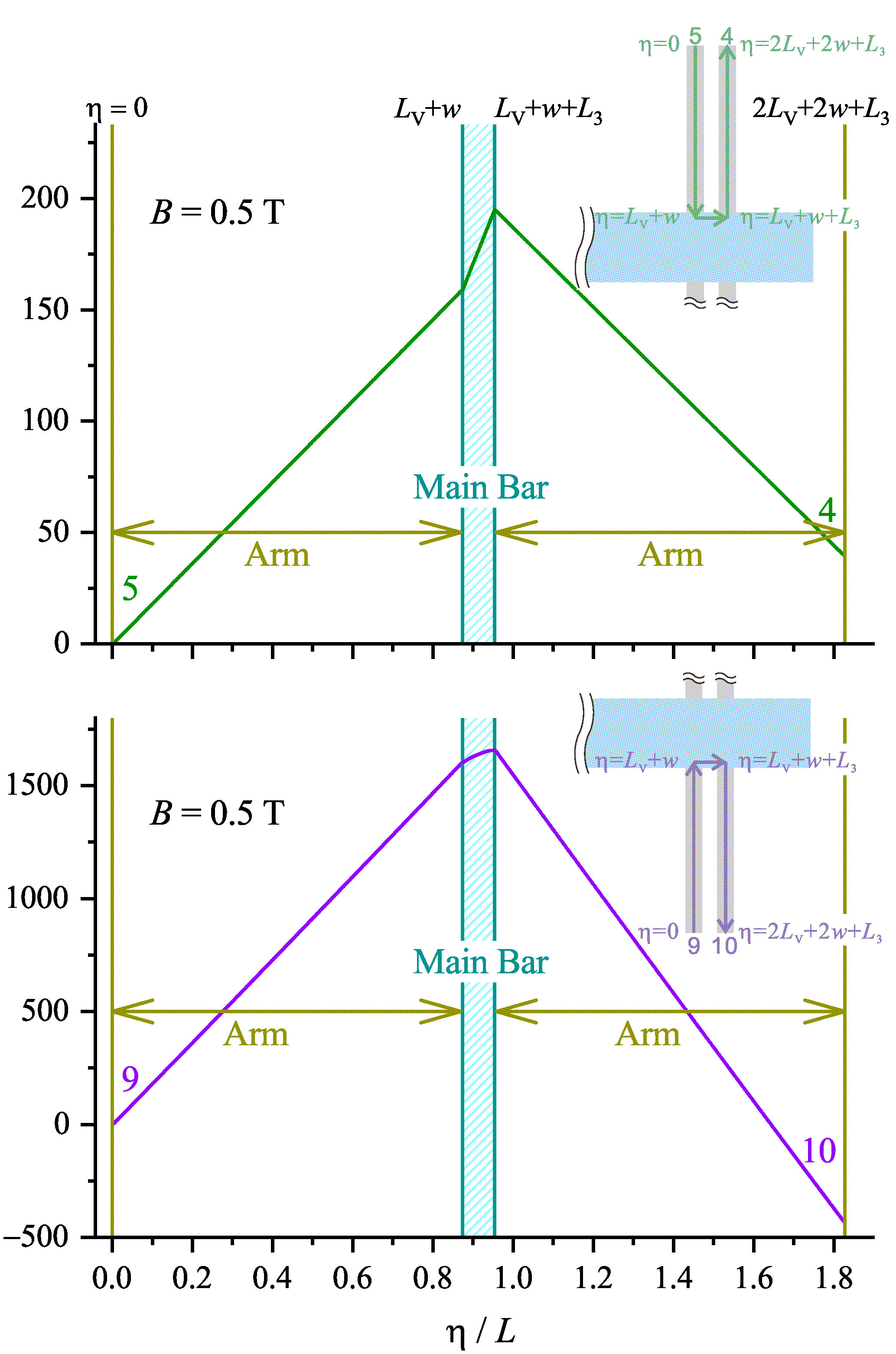} 
\caption{$\Delta \widetilde{\psi}_{xx}(\widetilde{\eta})$ given by equation (\ref{psixx}) for contact pairs 4-5 (top) and 10-9 (bottom) at $B = 0.5$ T\@. Insets: Schematic drawings depicting the integration path.}
\label{Integxx}
\end{center}
\end{figure}

The thermoelectric voltage $\phi_{i,j}$ between the contacts $i$ and $j$, to be compared with the experimentally measured value, is obtained by integrating $-{\bf E}$ along the path $C_{j \rightarrow i}$ connecting the contacts,
\begin{eqnarray}
\phi_{i,j}  =  - \int_{C_{j \rightarrow i}}^{} {\!\!\!\! \bf E}  \cdot d{\bf l } 
 = -\left( {\int_{C_{j \rightarrow i}}^{} {\!\!\!\! E_x} dx + \int_{C_{j \rightarrow i}}^{} {\!\!\!\! E_y} dy} \right) \nonumber \\
 = -s_{xx} \int_{C_{j \rightarrow i}}^{} {\!\!\!\! \boldsymbol{\nabla} \psi}  \cdot d{\bf l }
-s_{yx} \left( \int_{C_{j \rightarrow i}}^{} {\!\! \frac{\partial \psi}{\partial x}} dy - \int_{C_{j \rightarrow i}}^{} {\!\! \frac{\partial \psi}{\partial y}} dx \right). \nonumber \\ \label{phiij}
\end{eqnarray}
In our boundary conditions, we assumed that the temperatures of all the voltage contacts are the same as the temperature of the mixing chamber $T_\mathrm{L} = T_\mathrm{bath}$, and thus the first term in equation (\ref{phiij}) vanishes for all the combinations of $i$ and $j$. This indicates that under the present boundary conditions and the assumptions of the temperature-independent isotropic $\hat{\sigma}$, primarily resulting from our experimental conditions that the sample is immersed in the low-temperature $^3$He/$^4$He mixture, both $V_{xx}$ and $V_{yx}$ probe only the off-diagonal (Nernst) component $S_{yx}$ of the thermopower tensor and are insensitive to the diagonal (Seebeck) component $S_{xx}$. Note that the Nernst effect manifests itself in a quite counterintuitive manner owing to the deflection of the temperature gradient described in section \ref{calcT}. We therefore only have to calculate the integration in the second term, which can be rewritten as
\begin{equation}
\phi_{i,j} = -s_{yx} {T_\mathrm{L}}^2 \Delta \widetilde{\psi}_{i,j}, \label{phiijc}
\end{equation}
using a dimensionless integral
\begin{equation}
\Delta \widetilde{\psi}_{i,j}  =   \int_{C_{j \rightarrow i}}^{} {\!\! \frac{\partial \widetilde{\psi}}{\partial \widetilde{x}}} d\widetilde{y} - \int_{C_{j \rightarrow i}}^{} {\!\! \frac{\partial \widetilde{\psi}}{\partial \widetilde{y}}} d\widetilde{x}. \label{Deltapsiij}
\end{equation}
To be more specific, we calculated the path through the centre of the arms,
\begin{equation}
\Delta \widetilde{\psi}_{i,j} = \Delta \widetilde{\psi}_{yx}(\widetilde{W}+\widetilde{L}_\mathrm{V}), \label{Dpsiyx}
\end{equation}
corresponding to $V_{yx}$, and the path passing through the centres of the arms before and after going along the small distance $w = 1$ $\mu$m from the edge of the main bar (see the insets to figure \ref{Integxx}),
\begin{equation}
\Delta \widetilde{\psi}_{i,j} = \Delta \widetilde{\psi}_{xx}(2\widetilde{L}_\mathrm{V}+2\widetilde{w}+\widetilde{L}_3), \label{Dpsixx}
\end{equation}
to be compared with $V_{xx}$.
In equations (\ref{Dpsiyx}) and (\ref{Dpsixx}), we used the integrations along the path, $\Delta \widetilde{\psi}_{yx}(\widetilde{y})$ and $\Delta \widetilde{\psi}_{xx}(\widetilde{\eta})$, defined as (see also figures \ref{HallBarFig}, \ref{Integyx} and \ref{Integxx}),
\begin{eqnarray}
\Delta \widetilde{\psi}_{yx}(\widetilde{y}) \equiv \int_{-\widetilde{W}-\widetilde{L}_\mathrm{V}}^{\widetilde{y}}\left. \frac{\partial \widetilde{\psi}(\widetilde{y}^\prime)}{\partial \widetilde{x}}   \right|_{x = x_n} {\!\!\!\!\! d\widetilde{y}}^\prime \hspace{4mm} (n = 1, 2, 3) \label{psiyx} \\
\hspace{25mm} (-W-L_\mathrm{V} < y < W+L_\mathrm{V}), \nonumber
\end{eqnarray}
where $n =1$, 2 and 3 are for the contact pairs 6-8, 5-9 and 4-10, respectively, and
\begin{eqnarray}
\Delta \widetilde{\psi}_{xx}(\widetilde{\eta}) \equiv \nonumber \\
\left\{ {\begin{array}{*{20}{l}}
\displaystyle \int_{\widetilde{y}_0}^{\widetilde{\eta}_1}  \left. \frac{\partial \widetilde{\psi}}{\partial \widetilde{x}} \right|_{x=x_2} {\!\!\!\!\! d\widetilde{y}} 
\hspace{15mm} (0 \leq \eta < L_\mathrm{V}+w) \\
\displaystyle \int_{\widetilde{y}_0}^{\widetilde{y}_1}  \left. \frac{\partial \widetilde{\psi}}{\partial \widetilde{x}} \right|_{x=x_2} {\!\!\!\!\! d\widetilde{y}} - 
\int_{\widetilde{x}_2}^{\widetilde{x}_2 + \widetilde{\eta}_2} \left. \frac{\partial \widetilde{\psi}}{\partial \widetilde{y}} \right|_{y = y_1} {\!\!\!\!\! d\widetilde{x}} \\
\hspace{25mm} (L_\mathrm{V}+w \leq \eta < L_\mathrm{V}+w+L_3) \\
\displaystyle \int_{\widetilde{y}_0}^{\widetilde{y}_1}  \left. \frac{\partial \widetilde{\psi}}{\partial \widetilde{x}} \right|_{x=x_2} {\!\!\!\!\! d\widetilde{y}}  - 
\int_{\widetilde{x}_2}^{\widetilde{x}_3} \left. \frac{\partial \widetilde{\psi}}{\partial \widetilde{y}} \right|_{y = y_1} {\!\!\!\!\! d\widetilde{x}}  + 
\int_{\widetilde{y}_1}^{\widetilde{\eta}_3}  \left. \frac{\partial \widetilde{\psi}}{\partial \widetilde{x}} \right|_{x=x_3} {\!\!\!\!\! d\widetilde{y}}, \\
\hspace{20mm} (L_\mathrm{V}+w+L_3  \leq \eta < 2 L_\mathrm{V}+2 w+L_3)
\end{array}} \right. \label{psixx}
\end{eqnarray}
with $x_1 \equiv L_1$, $x_2 \equiv L_1 + L_2$, $x_3 \equiv L_1 + L_2 + L_3$, $y_0 \equiv \pm (W+L_\mathrm{V})$, $y_1 \equiv \pm (W - w)$, 
$\eta_1 \equiv y_0 \mp \eta$, $\eta_2 \equiv \eta - (L_\mathrm{V} + w)$, $\eta_3 \equiv y_1 \pm (\eta - L_\mathrm{V} -  w - L_3)$, where the upper (lower) sign in $y_0$, $y_1$, $\eta_1$ and $\eta_3$ is for the contact pair 4-5 (9-10).

In the top panels of figures \ref{SimVyx} and \ref{SimVxx}, we plot $\Delta \widetilde{\psi}_{i,j}$ calculated with equations (\ref{Dpsiyx}) and (\ref{Dpsixx}), respectively, at $B = 0.0$, 0.1,..., 1.0 T for the contact pairs $(i,j)$ noted in the figure. The values of $\Delta \widetilde{\psi}_{i,j}$ for $(i,j) = (4,10), (5,9), (6,8), (4,5)$ and $-\Delta \widetilde{\psi}_{10,9}$ exhibit similar $B$-dependence: after initial steep rise from $B = 0.0$ T, they take a mild peak at $B \sim 0.4$ T, followed by a gentle decline. The behaviour basically tracks the $B$-dependence of the Hall angle $\delta$ (see figure S3 in the Supplementary data). This can be qualitatively understood by noting that $\delta$ plays a key role in determining the orientation of %the temperature gradient 
$\boldsymbol{\nabla} T$. The angle between $\boldsymbol{\nabla} T$ and the side edges generally increases with $\delta$, leading to the enhancement of $|\partial \widetilde{\psi} / \partial \widetilde{x}|$ in the arms and $|\partial \widetilde{\psi} / \partial \widetilde{y}|$ in the main bar, both resulting in the increase of $|\Delta \widetilde{\psi}_{i,j}|$. We note in passing that $\delta$ shows non-monotonic $B$-dependence because we employed $\sigma_\mathrm{UE}$ instead of semiclassical $\sigma_{xx} = n_\mathrm{e} e \mu / (1 + \mu^2 B^2)$ in calculating $\delta$. 

We can see the relations $\Delta \widetilde{\psi}_{6,8} < \Delta \widetilde{\psi}_{5,9} < \Delta \widetilde{\psi}_{4,10}$ and  $\Delta \widetilde{\psi}_{10,9} < 0 < \Delta \widetilde{\psi}_{4,5}$ with $|\Delta \widetilde{\psi}_{10,9}| \gg |\Delta \widetilde{\psi}_{4,5}|$ in figures \ref{SimVyx}  and \ref{SimVxx}, respectively. Using an interpolating spline curve connecting the data points, we calculate $\phi_{i,j}$ with equation  (\ref{phiijc}) to simulate the experimentally observed $V_{yx}$ and $V_{xx}$. The calculated $\phi_{i,j}$, plotted in the bottom panels of figures \ref{SimVyx} and \ref{SimVxx}, qualitatively reproduce essential features in the experimental traces shown in figures \ref{MeasVyx} and \ref{MeasVxx}, respectively. The variation of  the oscillation amplitude with the magnetic field is in agreement with the experiments in both figures. Diminishing amplitude with the distance %of the voltage contacts 
from the heater seen in $V_{yx}$ and the relation of the sign (phase) and the amplitude between the top and the bottom voltage contacts in $V_{xx}$ are also well reproduced. All of these traits inherit the behaviour of the  $\Delta \widetilde{\psi}_{i,j}$ mentioned above. By inverting the magnetic field, the sign of $\delta$, hence the orientation of $\boldsymbol{\nabla} T$ with respect to ${\bf j}_\mathrm{Q}$, is also inverted. The spatial distribution of the temperature can thus be obtained by simply mirroring those shown in figure \ref{TAephcalc} with respect to the $x$ axis ($y = 0$). Therefore the voltage between the top contacts (4-5) and the bottom contacts (10-9) simply change places, namely, the top contacts at $B < 0$ take the place of the bottom contacts at $B > 0$, and vice versa. This is basically in agreement with what we observe in the experiments (figure \ref{MeasVxx}).

To see which part of the device has made the dominant contribution to the thermoelectric voltages, we examine how the integral in equation (\ref{Deltapsiij}) varies during the course of the integration. In figure \ref{Integyx}, we plot $\Delta \widetilde{\psi}_{yx}(\widetilde{y})$ given by equation (\ref{psiyx}) as a function of $\widetilde{y}$ at $B = 0.0$ T and 0.5 T\@. Without the magnetic field, the contribution of the voltage arms is negligibly small and $\Delta \widetilde{\psi}_{i,j}$ almost exclusively derives from the main bar. Note, however, that the $\Delta \widetilde{\psi}_{i,j}$ do not yield a thermoelectric voltage, since $s_{yx} = 0$ at $B = 0.0$ T\@. In the magnetic field, by contrast, the voltage arms accounts for most of the integral. $\Delta \widetilde{\psi}_{i,j}$ $(>0)$ is primarily determined by the gain at the bottom voltage arm, which far exceeds the loss at the top voltage arm. The integral over the main bar is negligibly small since $\boldsymbol{\nabla} \psi$ is nearly parallel to $y$ direction there. Again by applying the analytic rectangular model to the arms, we can readily find that the integral over an arm increases with the temperature difference between the contact (at $T_\mathrm{L}$) and the edge of the main bar to which the arm is attached. Note that the gradient $|\partial \psi / \partial x|$ (corresponding to $\tau_y$ in equation (\ref{tauy})) is enhanced by the temperature difference through $\tau_0$ in equation (\ref{tau0}). The bottom arm dominates $\Delta \widetilde{\psi}_{i,j}$, since it is attached to the higher-temperature edge of the main bar.  The decrease of $\Delta \widetilde{\psi}_{i,j}$ with the distance from the heater traces the decrease in the temperature of the bottom edge, which, in turn, mainly results from the thermal diffusion through the arms into the contacts (see the thick solid red line in figure \ref{CSB05}).

Figure \ref{Integxx} shows $\Delta \widetilde{\psi}_{xx}(\widetilde{\eta})$ given by equation (\ref{psixx}) as a function of $\widetilde{\eta}$ at $B = 0.5$ T for the top and the bottom contact pairs. Again, the voltage arms dominate the integrals, although the main bar (the second term in equation (\ref{psixx}) containing $|\partial \psi / \partial y|$) also makes a discernible contribution. %owing to relatively large $-\partial \widetilde{\psi} / \partial \widetilde{y}$. 
For the top pairs, the voltage arms are attached to the locations of the main bar having nearly the same temperature (see the thick solid blue line in figure \ref{CSB05}). Therefore the contributions of the two arms cancel each other, and the relatively small integral from the main bar survives, leading to a small value of $\Delta \widetilde{\psi}_{i,j} > 0$. For the bottom pairs, contribution of the arms is much larger than the top pairs. Near the bottom edge of the main bar, the locations for the two voltage arms have a relatively large temperature difference, as can be perceived from the thick solid red line in figure \ref{CSB05}. Thus, the loss due to the arm of the voltage probe 10 far exceeds the combined gain by the voltage probe 9 and the main bar, resulting in a large negative value of $\Delta \widetilde{\psi}_{i,j}$. %The behaviour of $\Delta \widetilde{\psi}_{i,j}$ described here provides a good qualitative explanation of the relative amplitude and the sign of the thermoelectric voltages measured with the top and the bottom contact pairs.

\section{Discussion \label{discussion}}
\subsection{Possible sources of the discrepancy between measured and calculated thermoelectric voltages \label{discMandC}}
So far, we have shown that our calculations qualitatively reproduce the essential features of our measurements. Quantitatively, however, the calculated thermoelectric voltages are larger than the measured values roughly by a factor of 2.5 to 10. We presume that the discrepancy is mainly caused by our use of oversimplified boundary conditions to make the calculations tractable. First, we simply assumed that the higher-temperature end of the main Hall bar ($x = L$ and $-W/2 < y < W/2$) has a constant temperature $T_\mathrm{H}$ inferred from the SdH amplitude measured with the voltage probes located on the other side of the secondary (heater) Hall bar (see figure \ref{HallBarFig}). We have neglected the spatial variation of the temperature within the heater section, which can be especially pronounced when the magnetic field is applied and a hot spot is generated \cite{Hirayama13}. Second, we set all the Ohmic contacts at the temperature $T_\mathrm{L} = T_\mathrm{bath}$ of the mixing chamber in which the sample is immersed. The contacts can, however, have the temperature higher than the bath owing to the Kapitza resistance \cite{Lounasmaa74} between the helium and the metallic (AuGeNi) film constituting the contacts. The effect will be more apparent for the voltage contacts having a smaller area hence a higher thermal resistance, especially for those attached to the higher temperature edge of the main Hall bar (contacts 10, 9, 8) and thus carrying higher thermal flux. The possible increase of the temperature in these contacts can, in principle, reduce the resulting thermoelectric voltages. It is difficult, however, to quantitatively assess the decrement mainly owing to the difficulty in determining the area of the interface between the helium and the granular metallic film. 
Note, however, that the temperature difference between voltage contacts, if present, allows for the longitudinal component, $S_{xx}$, to contribute to the thermoelectric voltages by resuming the first term in equation (\ref{phiij}). As pointed out earlier, the footprint of $S_{xx}$ is barely discernible in the lineshape of the experimentally observed oscillations shown in figures \ref{MeasVyx} and \ref{MeasVxx}, suggesting the absence of appreciable temperature difference between the measured contact pairs. We therefore surmise that the temperatures of the voltage contacts are sufficiently close to $T_\mathrm{L}$.

We can also find possible sources of the quantitative disagreement between the calculations and the measurements apart from the boundary conditions. We tacitly assumed that the 2DES in the voltage arms have the same properties as those of the main bar. It is possible, however, that the quality of the 2DES has declined during the process of fabricating the narrow (3 $\mu$m) arms. This can alter the temperature distribution by reducing the amplitude of SdH oscillations hence $\sigma_{xx} = \sigma_\mathrm{UE}$, which determines $\delta$, $\gamma_\mathrm{D}$ and $\gamma_\mathrm{P}$. Probably more importantly, the amplitude of the oscillations in $s_{yx}$ can also be reduced, resulting in the reduction of the oscillation amplitude of the thermoelectric voltage generated at the arms. Neglecting the quantum Hall edge states can also be a source, especially in the higher magnetic field regime. All the effects described here are rather difficult to be unambiguously quantified and cannot readily be incorporated in our model. Despite the quantitative discrepancy, we believe that our simple model captures the essence of the spatial distribution of the temperatures and the resulting thermoelectric voltages in a Hall bar placed in a magnetic field.

\subsection{Characteristics and limitations of the present measurement configurations \label{discCharLim}}
As we have shown in the previous sections, thermoelectric voltages can be a good tool to probe the spatial distribution of $T$ with the aid of the known thermoelectric coefficients. To this end, we used thermoelectric coefficients deduced from the measured resistivity employing the generalized Mott's formula. In many cases, however, the purpose of measuring thermoelectric voltages is to extract unknown thermoelectric coefficients. Our present settings are obviously inappropriate for this purpose. As mentioned in section \ref{thermvol}, the measurement is insensitive to $S_{xx}$. Furthermore, voltages generated at the voltage-probe arms exceed those generated at the main bar, the presupposed arena for the measurements. The situation cannot be improved by making the arms shorter, since the effect of the shortening the integration path in equation (\ref{Deltapsiij}) is cancelled out by the increase in the gradient (see equation (\ref{tau0})). These drawbacks mainly stem from our experimental configuration in which the voltage contact pads are thermally in contact with the surrounding helium bath. In a standard experimental setup for thermoelectric measurements, a sample is placed in a vacuum and is cooled down by thermally connecting one end (the heat sink) to the refrigerant, and the temperature gradient is introduced by heating the other end \cite{Fletcher99,White79}. By selecting a material with a high thermal resistance for the wiring to the voltage contacts \cite{Fletcher86}, the contribution of the voltage arms can be reduced. It will still be difficult, however, to completely eliminate or precisely determine the temperature gradient at the voltage arms and the resulting thermoelectric voltages. Regardless of the contribution of the arms, the redistribution of the temperature within the main bar by the magnetic field should be taken into consideration in interpreting the measured thermoelectric voltages.  It is also worth pointing out that the electron temperature of GaAs/AlGaAs 2DES is quite difficult to cool down to the lowest temperature in a dilution refrigerator in this standard configuration (see, e.g., \cite{Samkharadze11}).

The magnetic field does not affect the thermal flux carried by phonons. Therefore, the effect of the magnetic field on the spatial distribution of the temperature becomes less significant in the measurement %carried out in the standard experimental setup 
employing an external heater to introduce temperature gradient also to the substrate, provided that the temperature is high enough for the thermal flux to be principally carried by phonons, and also for the electron-phonon coupling to be strong enough so that the electron system has the same temperature as the lattice system. In such circumstances, the thermoelectric power is dominated by the phonon-drag contribution \cite{Fletcher86}.

\section{Conclusions \label{conclusion}}
We have examined the effect of a magnetic field on the spatial distribution of the electron temperature and the resulting thermoelectric voltages generated in a 2DES residing in a Hall bar device fabricated from a GaAs/AlGaAs wafer. At the temperatures investigated in the present study ($T < T_\mathrm{H} =330$ mK), electrons and phonons are virtually decoupled, and the relations between the gradient $\boldsymbol{\nabla} T$  of the electron temperature (or more precisely, ${\bm{\tau}} = -\boldsymbol{\nabla} \psi$), the thermal flux density ${\bf{j}}_\mathrm{Q}$ and the source-drain temperature difference (or, difference in $\psi$) basically duplicate those between the electric field, the electric current density and the source-drain bias in the magnetic field. The temperature difference engendered between the top and the bottom edges of the main bar, resulting from the deflection of $\boldsymbol{\nabla} T$ from the source-drain direction by a large angle $\delta \sim \pi/2$, is most directly detected by the difference in the thermoelectric voltages $V_{xx}$ measured along the top and the bottom edges (figure \ref{MeasVxx}). The top and the bottom edges exchange their roles by inverting the magnetic field, owing to the sign reversal in $\delta$. A notable feature unique to the thermoelectric-voltage measurements is the significant roles played by the voltage arms. A magnetic field lets the thermal flux flow through the arms regardless of the thickness and the length, disabling the design of the low aspect ratio to suppress the thermal flux. The resultant drop in the temperature, as well as the substantial thermoelectric voltage generated within the arms themselves, is responsible for the decrease in the measured thermoelectric voltages $V_{yx}$ with the distance from the heater (figure \ref{MeasVyx}). The effects of the arms also enhance the difference in $V_{xx}$ between the top and the bottom edges. The redistribution and reorientation of the temperature by a magnetic field demonstrated in the present study pose a caveat to be borne in mind in interpreting the thermoelectric-voltage of a 2DES measured in a magnetic field at very low temperatures.

\ack
We acknowledge T.\ Okamoto for calling our attention to the problem with spatially varying $V_{yx}$. This work was supported by JSPS KAKENHI Grant Numbers JP26400311, JP17K05491.

\appendix
\setcounter{section}{1}
\section*{Appendix \label{PDPP}}
In this appendix, we derive equations (\ref{Pdef}) and (\ref{Ppz}) in the main text. The two contributions ($r =$ ``def'' or ``pz'') are given as \cite{Price82,KajiokaCO13}
\begin{equation}
P_r(T) = \Pi_r(T) - \Pi_r(T_\mathrm{L}), \label{Prsub}
\end{equation}
where
\begin{equation}
\Pi_\mathrm{def}(T) = \frac{D^2 {m^*}^2 v_\mathrm{l} {{a_\mathrm{B}}^*}^2}{16\sqrt{2} \pi ^{5/2} \hbar ^2 \rho} \left( \frac{k_\mathrm{B} T}{\hbar v_\mathrm{l}} \right)^7 G_\mathrm{l}^\mathrm{def}(\kappa_\mathrm{F}) \label{Pidef}
\end{equation}
and
\begin{equation}
\Pi_\mathrm{pz}(T) = \Pi_{\mathrm{pz},\mathrm{l}}(T) + 2  \Pi_{\mathrm{pz},\mathrm{t}}(T) \label{Pipztot}
\end{equation}
with
\begin{equation}
\Pi_{\mathrm{pz},s}(T) = \frac{\left( e h_{14} \right)^2 {m^*}^2 v_s {{a_\mathrm{B}}^*}^2}{16 \sqrt{2} \pi ^{5/2} \hbar ^2 \rho} \left( \frac{k_\mathrm{B} T}{\hbar v_s} \right)^5 G_s^\mathrm{pz}(\kappa_\mathrm{F}), \label{Pipz}
\end{equation}
and $s = $ l and t represent longitudinal and transverse modes, respectively. (Strictly speaking, these formulae are for $B = 0.0$ T and do not take the Landau quantization into consideration. We assume that they are still applicable in the relatively weak magnetic field regime studied in the present paper \cite{KajiokaCO13}.) %, where the Landau quantization manifests itself as oscillatory modulation in the density of states, rather than descrete quantized levels, owing to considerable overlap between adjacent disorder-broadened Landau levels.) 
We used the following material parameters for GaAs \cite{Adachi85,Lyo88,Lyo89,Endo05HH,KajiokaCO13}: the deformation potential $D = -9.3$ eV, the piezoelectric constant $h_{14} = 1.2\times10^9$ V$\cdot$m$^{-1}$, the mass density $\rho = 5.3$ g$\cdot$cm$^{-3}$, the effective Bohr radius $a_\mathrm{B}^* =10.4$ nm, the effective mass $m^* =0.067m_e$ (with $m_e$ the bare electron mass) and the longitudinal and transverse sound velocities $v_\mathrm{l} = 5.14\times10^3$ m$\cdot$s$^{-1}$ and $v_\mathrm{t} = 3.04\times10^3$ m$\cdot$s$^{-1}$, respectively.
The dimensionless function $G_s ^r (\kappa_\mathrm{F})$ is written as
\begin{eqnarray}
G_s ^r (\kappa_\mathrm{F}) \equiv &
\displaystyle{\frac{1}{\pi} \int_{-\infty}^\infty  d\zeta \left| F (q_{Ts} \zeta ) \right|^2 \times \hspace{25mm}} \nonumber \\
 & \!\!\!\!\!\!\!\! \displaystyle{\int_0^{\kappa_\mathrm{F}} \! \! \! \! \! \frac{d\xi}{\sqrt{1 - (\xi / \kappa_\mathrm{F})^2}} \frac{g_s^r (\xi , \zeta)}{e^{\sqrt{\xi ^2  + \zeta ^2 } }  - 1} \frac{1}{H^2( q_{Ts} \xi )} },
\label{Gsr}
\end{eqnarray}
using the form factor, 
\begin{equation}
F( q_z ) = \int {dz |\Phi (z)| ^2} e^{iq_z z},
\label{FormFactor}
\end{equation}
and a function,
\begin{equation}
H( q_{\|} ) = \int \!\!\!\! \int {dz_1 dz_2 |\Phi (z_1)| ^2  |\Phi (z_2)| ^2} e^{-q_{\|} |z_1 - z_2|},
\label{ScrnFactor}
\end{equation}
with $q_z$ and $q_\|$ the phonon wavevector perpendicular and parallel to the 2DES plane, respectively, and $\Phi (z)$ the envelope of the 2DES wavefunction in the $z$ direction. 
In equation (\ref{Gsr}), we used $q_{Ts} \equiv k_\mathrm{B} T / \hbar v_s$ to normalize 2$k_\mathrm{F}$ and the components of the phonon wavevector: $\kappa_\mathrm{F} \equiv 2 k_\mathrm{F}/q_{Ts}$, $\xi \equiv q_\|/q_{Ts}$, and $\zeta \equiv q_z/q_{Ts}$.
The kernel $g_s^r (\xi , \zeta)$ in the integral of equation (\ref{Gsr}) is given as \cite{Price82}
\begin{equation}
g_\mathrm{l} ^\mathrm{def} (\xi , \zeta) \equiv \xi ^2 (\xi ^2 + \zeta ^2) ^{3/2},
\label{gdefl}
\end{equation}
\begin{equation}
g_\mathrm{l} ^\mathrm{pz} (\xi , \zeta) \equiv 
\frac{ 9 \xi ^6 \zeta ^2 }{2 (\xi ^2 + \zeta ^2) ^{5/2} },
\label{gpzl}
\end{equation}
and
\begin{equation}
g_\mathrm{t} ^\mathrm{pz} (\xi , \zeta) \equiv 
\frac{ 8 \xi ^4 \zeta ^4 + \xi ^8}{4 (\xi ^2 + \zeta ^2) ^{5/2} }.
\label{gpzt}
\end{equation}
For the temperature range and $n_\mathrm{e}$ in the present study, $\kappa_\mathrm{F}$ becomes large enough for the values of $G_s ^r (\kappa_\mathrm{F})$ to be virtually independent of the temperature, $G_\mathrm{l} ^\mathrm{def} (\kappa_\mathrm{F}) \simeq 363$, $G_\mathrm{t} ^\mathrm{pz} (\kappa_\mathrm{F}) \simeq 4.37$ and $G_\mathrm{t} ^\mathrm{pz} (\kappa_\mathrm{F}) \simeq 2.87$. Substituting these numbers and GaAs material parameters described above into equations (\ref{Pidef}) and (\ref{Pipz}), we attain, from equation (\ref{Prsub}), equations (\ref{Pdef}) and (\ref{Ppz}) with the values of coefficients $P_\mathrm{D}$ and $P_\mathrm{P}$ noted just below the equations.

\bibliography{ourpps,lsls,thermo,twodeg,qhe}

\clearpage
{\large \bf Supplementary data}\\

\includegraphics[width=17cm,clip]{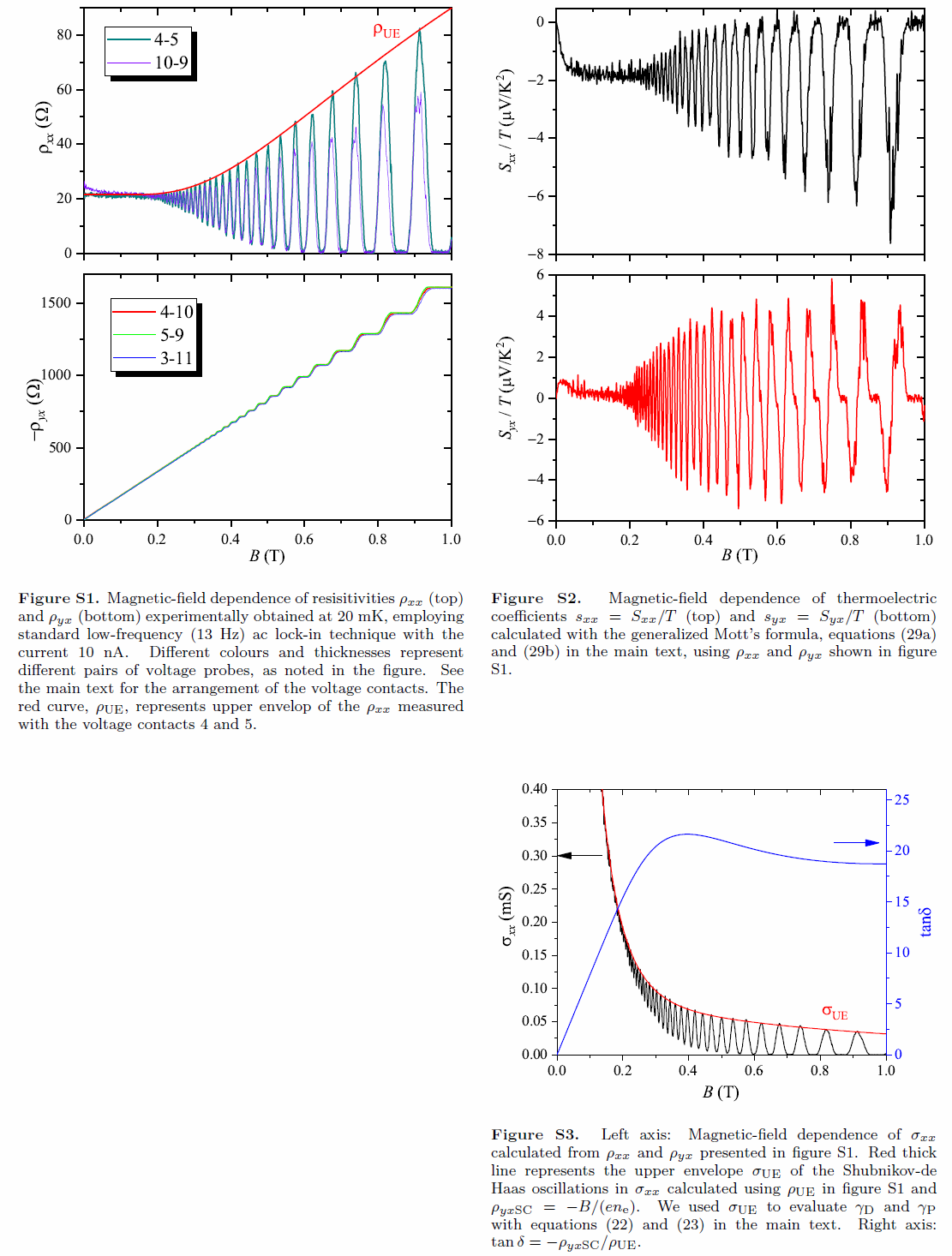}
\end{document}